\documentclass[11pt,superscriptaddress,eqsecnum,tightenlines,floats,aps,amsmath,
amssymb,nofootinbib,prd,noshowpacs,longbibliography]{revtex4}
\usepackage[utf8]{inputenc}
\usepackage{amsmath,amsthm,latexsym,amssymb,amsfonts,color,graphicx}
\usepackage{hyperref}

\def\be{\begin{equation}}
\def\ee{\end{equation}}
\def\ba{\begin{eqnarray}}
\def\ea{\end{eqnarray}}

\newcommand\nn{\nonumber}
\newcommand{\q}{\quad}

\newcommand{\im}{\mathrm{i}}

\renewcommand{\d}{{\mathrm d}}
\newcommand{\p}{{\partial}}

\renewcommand{\L}{{\mathcal L}}

\newcommand{\const}{{\text{const}}}

\begin{document}

\title{Holographic formulation of 3D metric gravity with finite boundaries }

\author{Seth K. Asante}\email{sasanteATperimeterinstitute.ca}
\affiliation{Perimeter Institute for Theoretical Physics, 31 Caroline Street North, Waterloo, ON, N2L 2Y5, CAN}
\affiliation{Department of Physics and Astronomy, University of Waterloo, 200 University Avenue West, Waterloo, ON, N2L 3G1, Canada}
\author{Bianca Dittrich}\email{bdittrichATperimeterinstitute.ca}
\affiliation{Perimeter Institute for Theoretical Physics, 31 Caroline Street North, Waterloo, ON, N2L 2Y5, CAN}
\author{Florian Hopfmueller}\email{fhopfmuellerATperimeterinstitute.ca}
\affiliation{Perimeter Institute for Theoretical Physics, 31 Caroline Street North, Waterloo, ON, N2L 2Y5, CAN}
\affiliation{Department of Physics and Astronomy, University of Waterloo, 200 University Avenue West, Waterloo, ON, N2L 3G1, Canada}

\begin{abstract}
In this work we construct holographic boundary theories for linearized 3D gravity, for a general family of finite or quasi-local boundaries. These boundary theories are directly derived from the dynamics of 3D gravity by computing the effective action for a geometric boundary observable, which measures the  geodesic length from a given boundary point to some centre in the bulk manifold. We identify the general form for these boundary theories and find that these are Liouville like with a coupling to the boundary Ricci  scalar.  This is illustrated with various examples, which each offer interesting insights into the structure of holographic boundary theories.
\end{abstract}

\maketitle

\tableofcontents

\section{Introduction}

Holographic dualities, for instance the AdS/CFT framework, suggest that a theory of quantum gravity can be dually described by a field theory defined on an asymptotic boundary.   That is, the partition functions of such dual boundary field theories, which would depend on the (asymptotic) boundary metric, can be interpreted as a partition function for gravity, which is however restricted to asymptotic boundary data.  

Here we are interested in extending such holographic dualities to finite and more general boundaries. One reason is that the partition function with boundary can also serve as  the vacuum (physical) wave function for gravity \cite{Oeckl}. Thus  aiming to employ holography to construct such physical wave functions, we need to understand such dualities for arbitrary boundaries.

A holographic boundary field theory would allow an easier access to the partition function of quantum gravity: instead of solving the full bulk dynamics of quantum gravity for given boundary data, and deal with the diffeomorphism gauge theory, one would have to ``just" solve the dynamics of the boundary field theory. For this to be a useful approach, the boundary field theory should be ideally local or an approximation to a local theory, with finitely many fields.  Note that otherwise the notion of holographic boundary field theory is quite empty, as one can construct boundary field theories by integrating out almost all bulk fields, except some degrees of freedom that one can attribute to the boundary. This will however generically lead to non--local boundary field theories, which could be converted to local ones at the price of introducing infinitely many fields.

The construction of ``quasi-local" holographic dualities has already been quite successful for 3D gravity. Here, due to the topological nature of the theory, one can indeed expect to encounter a local field theory, if one goes through the procedure described above. Thus there are a number of approaches in which such boundary field theories for gravity can be constructed. Moreover, again due to the fact that there are no propagating bulk degrees of freedom, the boundary field theories describe so-called boundary degrees of freedom, which in the case of gravity can often be understood as encoding the shape of the boundary in the embedding space time.

This starts with the relation between the Chern--Simons description of gravity \cite{3DGravChernSimons} to its WZW boundary theory \cite{WZW1,WZW2,WZW3}, which however relies on connection boundary data.  Restricting to  asymptotically AdS or asymptotically flat boundary conditions one obtains a Liouville (like) boundary field theory \cite{Lio1,Lio2,Henneaux,BarnichPhi,Maloney}. See also \cite{Bautier,Skenderis,CarlipBreakingDiffeoAdS} for  derivations of Liouville theory for the asymptotic AdS boundary which do not start from the Chern--Simons formulation.

Using metric boundary data \cite{BonzomDittrich} showed that a Liouville like dual field theory can also be identified more directly for finite boundaries. This work considered a specific background space time, so-called twisted thermal flat space \cite{TwistedFlatSpace}, and employed (linearized) Regge calculus \cite{Regge}, a discretization of gravity, in which the variables are given by edge lengths in a piecewise flat geometry. Thus  these variables can be identified as geodesic lengths. Using discretization independence of the one-loop partition function of the theory \cite{steinhaus11}, one can choose a discretization in which a class of variables describes the geodesic lengths from the boundary to some central axis.  These can be taken as boundary field variables, and one can thus easily integrate out  all variables except these boundary field variables.
 \cite{BonzomDittrich} also computed  the one-loop partition function for a finite boundary, which led to the same result as for asymptotically flat boundaries \cite{barnich1502}, see \cite{Giombi} for a corresponding result for asymptotic AdS boundaries. 

In the approach of \cite{BonzomDittrich}, the boundary field theory is directly derived from gravity and obtained as an effective action for a geometric observable, which encodes the shape of the boundary.\footnote{Note that this is not a Dirac observable, as Dirac observables should be independent of the shape of the boundary, and are very hard to come by, see e.g. \cite{DittrichObs1,DittrichObs2}. Here the chosen geometrical observable should rather encode the shape of the boundary.} Thus one has the advantage that the boundary field theory gives direct access to the dynamics of a geometric observable, which allows a ``bulk reconstruction". Such effective actions for geometric observables have also been studied independently from holographic considerations \cite{ReuterObs1,ReuterObs2}.  

The choice of the geodesic distance from the boundary to a centre also resonates with earlier studies \cite{CarlipBreakingDiffeoAdS}, which argued that boundary degrees of freedom arise in gravity due to the fact that the boundary breaks diffeomorphisms. In fact, we will see that the geodesic distance captures the change in the shape of the boundary that arises from diffeomorphisms generated by vector fields normal to the boundary.

Quasi-local holographic dualities have also been derived in a completely non-perturbative framework \cite{DGLR1,DGLR2,DGLR3,Riello4,Wolf18}, in particular for the Ponzano-Regge model \cite{PR} of 3D quantum gravity. This model constitutes a quantization of first order (Palatini) gravity. It offers precise control on the (quantum) boundary conditions and  their (quantum) geometric interpretations via loop quantum gravity techniques \cite{LQGTech}. In particular one can again choose (quantum) metric boundary conditions. Different kinds of boundary field theories arise, e.g. in the form of spin chain models, or in the form of sigma models, depending on the precise  choice of (quantum) boundary conditions and the choice of geometric variable that describes the embedding of the boundary. In particular \cite{Riello4} provides a fully non-perturbative version of having the geodesic lengths as a boundary field, in which case one obtains so--called RSOS models as boundary theories. \cite{DGLR2} performs the semi-classical analyses for a particular family of boundary conditions, which are encoded in a particular choice of boundary wave functions \cite{LivineSpeziale}. This led to a confirmation of the one-loop partition function found in \cite{barnich1502,BonzomDittrich}, albeit  with Planckian corrections, which arise due to the fact that the Ponzano--Regge framework  allows for an arbitrary winding number of the boundary around the central axis.

To a great extend these works rely on the topological nature of 3D gravity. Thus the question arises whether these constructions can be also applied to 4D gravity. A first step to answer this question can be found in \cite{ADH}, which uses again (linearized) Regge calculus to consider a background spacetime, which is a 4D version of twisted thermal flat space. Restricting to boundary data which induce a 4D flat solution, \cite{ADH} finds the same type of boundary theory as in 3D. However, due to the fact that 4D Regge calculus does not feature a local discretization independent measure \cite{DittrichKaminskiSteinhaus}, it is hard to extend this result to the (one-loop) quantum theory.\footnote{One can consider a  model for quantum flat space \cite{Baratin4D}, for which a discretization independent model does exist. In this case one can compute the one-loop partition function \cite{ADH}, which  captures the effect of the boundary degrees of freedom.}

To extend these results to more general backgrounds and to tackle the main task, namely including gravitons, we need a framework that is applicable to 4D gravity and for which we can expect to solve the dynamics. Being particularly interested in  length observables, we will therefore consider (linearized) metric gravity. As the geodesic lengths   has so far been shown to be a convenient choice for the boundary field, which moreover is connected to obtaining Liouville (like) boundary theories, we will stick with this choice. This does however 
 present us with a challenge, namely to compute the effective action for a composite observable. 
 
 In this work we will therefore go back to 3D gravity to develop and test a general framework in which such effective actions can be computed. As we will see this allows us to consider more general backgrounds and boundaries and to systematize and greatly extend  the results which have been obtained so far. 
 
 A first key result will be the computation of the  Hamilton--Jacobi functional for 3D (linearized) gravity for a large family of boundaries\footnote{We consider boundaries with homogeneous intrinsic  curvature $\partial_A \, {}^2\!R=0$ and with non-vanishing extrinsic curvature.}, in section \ref{BigSecHJ}. (This amounts to the classical limit of the physical vacuum wave function associated to the given boundary.) It turns out that a convenient way to express this Hamilton--Jacobi function is in terms of the diffeomorphism generating vector field that generates the on-shell metric perturbations. In fact, the Hamilton--Jacobi functional is local in terms of these vector fields. Note however that the vector fields themselves are non-local functionals of the boundary metric data. 

This allows us to propose  in section \ref{secDual} a  field theory for a scalar field defined on the boundary, whose Hamilton--Jacobi functional reproduces the one for gravity and whose equation of motion imposes that the scalar field equals the geodesic lengths from the boundary to some centre. The proposed action confirms the earlier findings of Liouville like boundary theories: for a boundary with background intrinsic metric $h_{AB}$ and background extrinsic curvature $K_{AB}$ the scalar field is governed by a quadratic form $\rho\Delta\rho:=\rho(2(K^{CD}-Kh^{CD})D_CD_D-{}^2\! R K)\rho$, where $\rho$ denotes the scalar field and $D_C$ denotes the covariant derivative compatible with $h_{AB}$. Additionally the scalar field has a Liouville coupling to the (first order perturbation of the) boundary Ricci scalar, that is the full  Lagrangian is given by  ${\cal L}= \sqrt{h} (\rho\Delta\rho-2\rho \delta({}^2\!R))$, where $\delta({}^2\!R)$  of the first order perturbation of the boundary Ricci scalar.


For this proposed boundary field theory we will however ignore some subtlety, which is the precise definition of the `centre' at which the geodesics starting from the boundary end. We will take care of this subtlety in the subsequent examples and will see that it can lead to a certain modification for the boundary field theory.

In the  sections \ref{SecFlat} to \ref{SecSph}, we will more directly compute the effective action for the geodesic lengths, using a Lagrange multiplier method, which we introduce in section \ref{BigSecEffGL}. We will see in section \ref{SecFlat}  and \ref{SecAdS} that a priori this method does {\it not} lead to the expected results for the cases of backgrounds with intrinsically flat boundaries, such as the torus boundaries appearing for the twisted thermal flat and AdS spaces, which form our first two examples.  The reason is that the geodesic lengths turns out to be in a certain sense a degenerate observable. This can be changed however by carefully implementing smoothness condition at the central axis of the solid torus. This procedure will lead to an effective action, which differs from the one proposed  in section \ref{secDual} by the insertion of a non-local operator. This insertion also implements a remnant of the diffeomorphism symmetry of the gravitational theory, which turns the precise location of the central axis into a gauge degree of freedom, also for the boundary field theory.

These findings confirm the boundary field theory found in \cite{BonzomDittrich} for the flat space example.\footnote{The smoothness conditions are there however automatically implemented with the (Regge) formalism.} We will also find that the one-loop partition functions of the boundary field theories for the twisted thermal flat and AdS spaces reproduce the gravitational one-loop partition functions \cite{Giombi,barnich1502,BonzomDittrich}.

The last example, which we consider in section \ref{SecSph}, is a spherical boundary in flat space (and thus with intrinsic background curvature), which has so far not been discussed in the literature. Here the mechanism for constructing the effective action differs slightly from the one with flat boundaries, as the smoothness conditions at the centre play less of a key role. The effective action will be local and agree with the proposed one from section \ref{secDual}.

We will close with a discussion and outlook in section \ref{SecDisc}. To avoid deviating from the key points in the main body of the paper, we deferred all more involved calculations and proofs to the appendices. This includes a summary of the conventions used in Appendix \ref{AppConv}, the defining formulas for a convenient parametrization of the perturbative boundary metric in Appendix \ref{AppB}, and the calculation of the 3D Hamilton--Jacobi functional in Appendix \ref{App2HJ}. Appendix \ref{AppD} evaluates the commutator of two specific operators ($\Delta$ and the radial derivative $\partial_r$, which here serves as a kind of time evolution operator), which is needed for the subsequent appendices \ref{AppE} and \ref{AppF}.  We discuss the derivation of solutions to the linearized Einstein equations with a Lagrange multiplier term in  Appendices \ref{AppE} and \ref{AppSph}. Appendix \ref{AppF} evaluates  the Lagrange multiplier dependent boundary term, Appendix \ref{AppG} construct the (linearized) geodesic length observable, and in Appendix \ref{AppSmooth} we derive the smoothness conditions which we need to implement at the centre of the bulk manifolds. Appendix \ref{AppEff} discusses the computation of effective actions for observables, which in a certain sense are degenerate. Finally, Appendix \ref{AppHarm} collects definitions for spherical vector and tensor harmonics, which are useful to discuss the example with spherical boundary in section \ref{SecSph}.

\section{The Hamilton--Jacobi functional for 3D gravity }\label{BigSecHJ}
 
 In this section we will determine the  Hamilton--Jacobi functional, that is the on-shell action, for 3D linearized gravity, for a large class of boundaries. To start with we will summarize our conventions and define the type of boundaries we will be considering. We will then introduce a convenient parametrization of the boundary metric perturbations in terms of diffeomorphism generating vector fields. The first key result we will present is to invert the relationship between metric perturbations and the vector field, that is to express the vector field components in terms of the boundary metric perturbations. We will then move on to our second key result, which is the evaluation of the Hamilton--Jacobi functional. This turns out to be a local functional, if we use the parametrization in terms of diffeomorphism generating vector fields.

 \subsection{Assumptions and Conventions}\label{secAssum}
 
 Here we consider 3D linearized gravity, with Euclidean signature, with or without a cosmological constant $\Lambda$, on a manifold $M$ with smooth boundary $\partial M$.  We consider vacuum 3D general relativity, that is all (background) solutions have homogeneous curvature
$ 
 R_{abcd} =  \Lambda \left( g_{ac} g_{bd} - g_{ad} g_{bc} \right) 
$.
 We choose for the background solution Gaussian coordinates
  \ba\label{bgmetric}
g_{ab}dx^a dx^b &=& dr^2 +  h_{AB} dy^A dy^B   \q ,
 \ea
 and assume that $r=0$ defines a  point or  a one--dimensional submanifold in $M$. We also assume that the boundary $\partial M$ is given by the set of points with fixed radial coordinate $r=r_{\rm b}$.  Here we denote with indices  $a,b, \ldots$ space--time indices and with $A,B,\ldots=1,2$  ``spatial" indices for the surfaces $r=\const$. We also use  $\perp$ as index for the radial coordinate.

 We will consider perturbations of the background metric 
 \ba
  g^{\text{full}}_{ab}&=&g_{ab} +\gamma_{ab}
 \ea
and describe with $\gamma_{\perp\perp}, \gamma_{\perp A}$ and $\gamma_{AB}$ the various components of the metric perturbations according to the foliation defined by the $r=\const$ surfaces. 

 For a two--dimensional (boundary) metric the Ricci  tensor is determined by the Ricci scalar $^2 \!R_{AB} = \frac12\ ^2 \!R h_{AB}$. We assume that the background boundary curvature is homogeneous $\p_A {}^2\! R = 0$.  In the next section \ref{secbasis} we will see that we will also need to assume a non--vanishing extrinsic curvature for the background boundary. More precisely we consider boundaries for which the relation between the boundary metric perturbations and the diffeomorphism inducing vector field leading to these boundary metric perturbations, is invertible.

With a Gaussian metric the Christoffel symbols are given by
 \ba\label{Chris1}
  \Gamma^a_{\perp\perp}  \,=\, 0  \, ,\q \Gamma^\perp_{\perp B}=0 \, ,\q   \Gamma^{\perp}_{AB}=-K_{AB} \, , \q \Gamma^A_{\perp B}=K^A_B \, ,\q \Gamma^A_{BC}= {}^2\!\Gamma^A_{BC}  \q ,
 \ea
 where the extrinsic curvature tensor is given by $K_{AB} =\tfrac{1}{2} \partial_\perp h_{AB}$.

 This allows to express the relations between space--time covariant derivatives and spatial covariant derivatives, e.g. 
 \ba
\nabla_A\xi_B &=&
D_A \xi_B  + K_{AB} \, \xi_\perp \q , \nn\\
\nabla_A \xi_\perp &=&  D_A(\xi_\perp)  - K^B_A \xi_B
\ea
where $\xi_\perp$ is treated as a spatial scalar, that is $D_A\xi_\perp=\partial_A \xi_\perp$. We use  $\nabla$ for the covariant derivative compatible with $g$, and $D$ for the covariant derivative compatible with $h$.


In  Appendix \ref{AppConv} we collect our conventions for the curvature tensors and  the  Gauss--Codazzi relations. We will in particular need that the Gauss--Codazzi relations imply for a surface embedded into a 3D vacuum solution
\ba\label{GCshort}
K^2 - K_{AB} K^{AB} \,=\, ^2\!R - 2\Lambda \q , \q\q  D_A {K^A}_B - D_B K =0 \q .
\ea

  \subsection{ A basis for the boundary metric perturbations }\label{secbasis}
  
  In 3D vacuum gravity the solutions to the equations of motion are diffeomorphism equivalent to a homogeneously curved space time. We can therefore express the metric perturbations in terms of the diffeomorphism generating vector fields.
  In fact, the Hamilton-Jacobi functional for 3D gravity will appear in a particular simple form, if we  parametrize the boundary metric perturbations via the diffeomorphism generating vector fields $\xi^a$. That is,  $\gamma_{AB}$ is parametrized in terms of the vector components $\xi^\perp$ and $\xi^A$ by  
 \ba\label{eq:gammaAB_in_xi}
 \gamma_{AB} \,=\, [ {\cal L}_{\xi} g ]_{AB} &=&    \nabla_A \xi_B + \nabla_B \xi_A  \nn\\
&=& 2\xi^\perp K_{AB}+ [{\cal L}_{\xi^\parallel} h]_{AB} \q .
 \ea
For this work we will assume that the transformation  from $(\xi^\perp, \xi^1,\xi^2)$  to $(\gamma_{11},\gamma_{22},\gamma_{12})$ is invertible. Clearly, this requires that the extrinsic curvature tensor $K_{AB}$ is non-vanishing. To explicitly invert this transformation requires some calculations, which we detail in Appendix \ref{AppB}. Here we state only the result:
 ~\\
{\bf Result 1:} The vector components $\xi^\perp$ and $\xi^A$ are determined by the equations
\ba\label{xiofg}
\Delta \, \xi^\perp &=& \Pi^{AB} \gamma_{AB} \nn\\
 {{\cal D}^A}_B \,  \xi^B &=&2(K^{BC}-K h^{BC}) \, \delta\, {}^2\! \Gamma^A_{BC}
\ea
where 
\ba
 \Delta &=&  2 (K^{CD} - K h^{CD}) D_C D_D - {}^2\! R \,K \q ,\nn\\
  {  {\cal D}^A}_B &=& \, 2 \left( K^{CD} -K h^{CD} \right)D_C D_D \,{h^A}_B - {}^2\! R {K^A}_B   \q ,   \nn\\
 \Pi^{AB} &=& D^A D^B - h^{AB} D_C D^C - \frac12 {}^2\! R\,h^{AB}  \q ,\nn\\
 \delta \,{}^2\! \Gamma^A_{BC}&=& \tfrac{1}{2} h^{AD} \left( D_B \gamma_{AC} +D_C \gamma_{BA} -D_A\gamma_{BC}\right) \q .
\ea
 
To obtain $\xi^\perp$ and $\xi^A$ we need to invert the operators $\Delta$ (on the space of spatial scalars) and ${{\cal D}^A}_B$ (on the space of spatial vectors). Thus the vector components are non-local functionals of the spatial metric perturbations.  Note that by construction, $\xi^\perp$ is a functional of the boundary metric perturbations, which is invariant under (linearized) boundary tangential diffeomorphisms. That is, $\Pi^{AB}$ is zero on perturbations induced by boundary tangential diffeomorphisms.  This suggest a relation of $\Pi^{AB}\gamma_{AB}$ to the first variation of the boundary Ricci scalar $\delta( {}^2\! R )$, which is also vanishing on boundary tangential diffeomorphisms . In fact,
\ba\label{2Ric}
 \Pi^{AB} \gamma_{AB}&=& (D^A D^B - h^{AB} D_C D^C)\gamma_{AB} - \tfrac{1}{2} {}^2\! R\,h^{AB} \gamma_{AB} \nn\\
 &=&(D^A D^B - h^{AB} D_C D^C)\gamma_{AB} -  {}^2\! R^{AB} \gamma_{AB} \nn\\
 &=& \delta( {}^2\! R ) \q ,
\ea
as we have ${}^2\!R_{AB}= \frac 12 \, {}^2\!R\, h_{AB}$ for two--dimensional metrics.

Having found the vector components $\xi^\perp$ and $\xi^A$ as functions of the boundary metric components, we can also express the lapse $\gamma_{\perp\perp}$ and shift $\gamma_{\perp A}$ of the metric perturbations as functions of the generating vector field $(\xi^\perp, \xi^A)$, and thus in terms of the boundary metric perturbations $\gamma_{AB}$:
\ba\label{SolLapShif}
\gamma_{\perp\perp} &=& 2 \partial_\perp \xi^\perp   \q , \nn\\
\gamma_{\perp A}&=& \nabla_\perp  \xi_A \,+\,  \nabla_A \xi_r \nn\\
&=& \partial_\perp (h_{AB} \xi^B) -    \Gamma_{\perp A}^B \xi_B  \,+\,  \partial_A (g_{\perp\perp}  \xi^\perp)   - \Gamma_{A\perp}^B \xi_B \nn\\
&=& D_A  \xi^\perp  \,+\,  h_{AB}\partial_\perp   \xi^B    \q .
\ea

  \subsection{The Hamilton--Jacobi functional}\label{secHJ}

   \subsubsection{Zeroth and first order contributions}

 The (Euclidean) Einstein--Hilbert action, with Gibbons-Hawking-York boundary term, is given by
\ba\label{actionC1}
S&=& -\frac{1}{2\kappa} \int_M d^{3} x \, \sqrt{g}  \left( R - 2 \Lambda \right) \,\,-\,\,\frac{1}{\kappa} \int_{\partial M} d^{2} y \sqrt{h}  \, \epsilon K   \q ,
\ea 
where $\kappa=8\pi G_N$ and $G_N$ is Newton's constant.  We will also use the following convention regarding the (sign of the) extrinsic curvature tensor: $K_{AB}$ will be understood as the extrinsic curvature tensor associated to the foliation of $M$ by surfaces of constant radius, that is, with our use of Gaussian coordinates, defined by $K_{AB}=\tfrac{1}{2}\partial_\perp h_{AB}$. This  differs however by a sign from the extrinsic curvature tensor associated to an inner boundary, which has 
outward pointing normal $n^a=(-1,0,0)$. We will therefore make the sign explicit and use the variable $\epsilon =\pm 1$. 

The equations of motions demand $R_{ab} = 2 \Lambda  g_{ab}$  and  thus  $R=6 \Lambda$. The action evaluated on a (background) solution is therefore given by
\ba
S&=&  -\frac{4\Lambda}{2\kappa}  {}^3\! V \,-\,\,\frac{1}{\kappa} \int_{\partial M} d^{2} y \sqrt{h} \epsilon K 
\ea
 where $ {}^3V$ is the volume of the manifold $M$.   
 
 The first variation of the action 
\ba\label{C17}
-\kappa\delta S &=&
\frac{1}{2}\int_M d^3 x \, \sqrt{g}\left( \left( \tfrac{1}{2}R -\Lambda \right) g^{ab} -R^{ab}\right) \delta g_{ab} \,+\,
\frac{1}{2}  \int_{\partial M} d^{2} y \sqrt{h} \epsilon \left(   K h^{AB}\  -  K^{AB}   \right)\delta g_{AB}\,  \q\q
\ea
 determines the (background) equations of motions as well as the first order of the on-shell action.  It also determines the momentum conjugated to the metric $\pi^{AB}= \sqrt{h}(K^{AB}-Kh^{AB})$.
 
 Using the parametrization $\delta g_{ab}=\gamma_{ab}= {\cal L}_\xi g_{ab}$ for the boundary metric fluctuations, the first order of the on-shell action evaluates to 
\ba
-\kappa S^{(1)}_{\rm HJ}
&=& \frac{1}{2}\int_{\partial M} d^{2} y 
\sqrt{h}\epsilon \left(   K h^{AB}  -  K^{AB}   \right) ( \nabla_{A} \xi_{B}  + \nabla_B \xi_A) \nn\\
&=& 
\int_{\partial M} d^{2} y 
\sqrt{h} \epsilon \left(   K h^{AB}\  -  K^{AB}   \right) \left( D_{A} \xi_{B} + K_{AB} \xi^\bot \right) \nn\\
&=&\int_{\partial M} d^{2} y 
\sqrt{h} \epsilon \left( \left(   -D^B K + D_A K^{AB} \right) \xi_B + \left( K^2 - K_{AB} K^{AB} \right) \xi^\bot \right) \nn\\
&=& \int_{\partial M} d^{2} y  \sqrt{h}\epsilon \;  \left( {}^{2}\!R -2\Lambda\right)\; \xi^\bot  \q .
\ea
where we have used the Gauss--Codazzi relations (\ref{GCshort}) to arrive at the last line.

Note that the presence of these first order terms may lead to second order terms which are {\it not} invariant under the tangential boundary diffeomorphisms, even if we have $ \left( {}^{2}\!R -2\Lambda\right)$=0. Such second order terms might be needed to make the full action invariant under tangential boundary diffeomorphisms to higher order.

  \subsubsection{Second order contributions}

 The second order bulk and boundary terms\footnote{The bulk and boundary terms are not uniquely determined, as one can redefine them using integration by parts. Here we have chosen a form, where the bulk term vanishes on-shell.}  of the action arise from the variation of the first order bulk and boundary terms respectively.  They are given by
 \ba\label{2oA01}
-\kappa  S^{(2)}
&=& 
\frac{1}{4} \int d^3 x  \sqrt{g} \, \,\gamma_{a b} \left(   
  V^{abcd} \,
  \gamma_{cd} \,\,+ \,\,
  \tfrac{1}{2}  \,
  G^{abcdef }
   \, \nabla_c \nabla_d \gamma_{ef}  \right) \,+ \nn\\
   &&
  \frac{1}{4} \int d^{2} y   \sqrt{h}\, \epsilon \,
\gamma_{ab}  \left(\,(B_1)^{abcd}\gamma_{cd} +   \, (B_2)^{abcde} \nabla_c \gamma_{de} \right)  \q ,
\ea
 where the tensors $V^{abcd}$ and $G^{abcdef }$ as well as $(B_1)^{abcd}$ and $(B_2)^{abcde}$ are detailed in Appendix \ref{AppE}. From the bulk term we can read of the equations of motion for the metric perturbations
 \ba\label{EOMLin}
  V^{abcd} \,
  \gamma_{cd} \,\,+ \,\,
  \tfrac{1}{2}  \,
  G^{abcdef }
   \, \nabla_c \nabla_d \gamma_{ef}  &=&0 \q .
 \ea
 Thus the second order contribution to the on-shell action comes only from the boundary term in (\ref{2oA01}). 3D gravity has no propagating degrees of freedom. This is due to the diffeomorphism symmetry, which renders the three degrees of freedom, given by the spatial metric perturbations $\gamma_{AB}$ to be gauge. This gauge symmetry also means that three of the six equations of motion in (\ref{EOMLin}) are redundant. With our assumptions, that  include that the transformation between the spatial metric perturbations $\gamma_{AB}$ and the diffeomorphism generating vector field $\xi^b$ are invertible, the three remaining equations of motions allow to determine the lapse and shift components  $\gamma_{\perp\perp}$ and $\gamma_{\perp A}$ as functions of the spatial metric $\gamma_{AB}$, see also equation (\ref{SolLapShif}). 

 Inserting these solutions for $\gamma_{\perp\perp}$ and $\gamma_{\perp A}$ into the boundary term in (\ref{2oA01}) one obtains a functional of the boundary metric fluctuations. Due to the fact that we do not have propagating degrees of freedom, this functional will not include any radial derivatives of $\gamma_{AB}$. In general we have however to expect that the Hamilton--Jacobi functional is boundary non-local, that is the integrand involves the inverse of (boundary) differential operators acting on $\gamma_{AB}$.

We detail in Appendix \ref{App2HJ} the evaluation of the boundary term, which shows that the  Hamilton--Jacobi functional is a local functional, if written in terms of the diffeomorphism generating vector field $\xi^b$.

~\\
{\bf Result 2:} 
The second order of the Hamilton--Jacobi functional is (with our assumptions stated in section \ref{secAssum}) given by 
 \ba\label{SHJ2}
 -\kappa  S^{(2)}_{\rm HJ}&=& \frac{1}{4} \int_{\partial M} d^2 y \sqrt{h} \epsilon \, \left(  \xi^\bot \Delta \, \xi^\bot \, -  \, \xi^A {\cal D}_{AB}  \xi^B  \right) \q ,
\ea
where 
\ba
\Delta&=& 2 (K^{CD} - K h^{CD}) D_C D_D - {}^2\! R \,K \q ,\nn\\
  {\cal D}_{AB} &=& \, 2 \left( K^{CD} -K h^{CD} \right)D_C D_D \,{h}_{AB} - {}^2\! R {K}_{AB} \q .
\ea

~\\
We see that the Hamilton--Jacobi functional  expressed in terms of the diffeomorphism generating vector field is has a strikingly simple form. Note  that the boundary normal component $\xi^\perp$ and the boundary tangential components $\xi^A$ of the diffeomorphism generating vector field decouple. This seems to hold specifically only in 3D gravity. The part invariant under the boundary tangential diffeomorphism is given by $\xi^\bot \Delta \, \xi^\bot = \delta({}^2\! R)  \Delta^{-1} \delta({}^2\!R)$. 

As we will see the lengths of geodesics which are normal (in the background geometry) to the boundary will be basically given by $\xi^\perp$. The differential operator $\Delta$ will therefore also be a key ingredient in the effective action for the geodesic lengths.

\section{Dual boundary field theories}\label{secDual}

Although the on-shell action is local as a functional of the $\xi^a$, it is a rather non-local functional of the boundary metric perturbations itself.  Here we are interested in defining a (local)  field theory, defined on the boundary $\partial M$, whose Hamilton--Jacobi functional agrees with the one of gravity.  We will refer to such a field theory as  dual boundary field theory.  

Moreover we would like to have boundary fields which can be identified with observables of the gravitational theory.  As the Hamilton--Jacobi--functional measures in particular the extrinsic curvature of the boundary, it is reasonable to look for observables which describe the shape of the boundary, or in other words, the embedding of the boundary in the (homogeneously curved) bulk solution. 

One such observable is the geodesic distance of a boundary point to a central bulk axis or a central bulk point at $r=0$.  More precisely we consider a geodesic from the point $(r_{\rm b},y^A)$  on the boundary $\partial M$ to the point $(r=0,y^A)$.  We can therefore understand the geodesic length as a field defined on the boundary itself.

Since the metric is of Gaussian form with respect to the radius and the boundary is a $r=\text{const.}$ surface, the tangent vector to  the geodesic is orthogonal to the boundary.  For this reason the geodesic length will be to first order in the (boundary) metric perturbations  invariant under boundary tangential diffeomorphisms.  Thus we can only expect to reproduce the part of the gravitational Hamilton--Jacobi functional, which is invariant under these boundary tangential diffeomorphisms, that is the part quadratic in $\xi^\perp$.   On the other hand, knowing that the first order of the geodesic lengths is boundary diffeomorphism invariant, we can suspect that it is proportional to $\xi^\perp$ evaluated on the boundary, which in turn is related to the first variation of the boundary Ricci scalar.

In the following we will determine the (first order of the) geodesic length as a function of the boundary metric. This will allow us to `guess' a candidate for a dual field theory, which (a) reproduces the equation of motion for this geodesic length and (b) reproduces the boundary diffeomorphism invariant part of the gravitational Hamilton--Jacobi functional. In the process we will encounter a subtlety, namely that the geodesic lengths is also affected by the position of the central axis or point. This position is determined by the bulk metric perturbations, which are however gauge degrees of freedom. 

The positions of a central point or axis do however only require three degrees of freedom for a central point and three degrees of freedom per axis point, whereas the boundary field describes one degree of freedom per boundary surface point. Indeed, we will  see later, that this arbitrariness  affects only certain momentum modes of the boundary field. But this feature will be also responsible for a certain modifications which arise, if we determine the action for the geodesic length more directly from the gravitational action.

In section \ref{secTang} we will furthermore find dual fields which reproduce the parts of the  gravitational Hamilton--Jacobi functional which describe the tangential boundary diffeomorphisms.

 \subsection{Action for the geodesic length}\label{secGeod}

 To start with we need to know the  lengths of geodesics $(r(\tau), y^A)$  as a functional of the metric perturbations to first order. As a second step we should express such geodesics as functionals of the boundary metric. 
 
 Note that the parametrized curves  $x^a(\tau)\,=\, (r_1+ (r_{\rm out}-r_{\rm in}) \tau,0,0)$ with $\tau \in [0,1]$  are  affinely parametrized geodesics with respect to  background metrics of the form (\ref{bgmetric}). This follows from the geodesic equation
 \ba
\frac{d x^a }{d\tau} \, \nabla_a \frac{d x^b}{d\tau}  \,=\,     \Gamma^b_{\perp\perp}  \, (r_2-r_1)^2 \,=\, 0 \q .
 \ea

 We now consider a geodesic $z^a(\tau)$ with respect to the full metric  $g^{\text{full}}_{ab}$ with fixed endpoints $z^a(0)$ and $z^a(1)$. As explained in Appendix  \ref{AppG}   its length is given to first order in  metric perturbations by
 \ba
 \ell \,=\, \frac{1}{2 (r_2-r_1)} \int_0^1 d\tau \,   \frac{d x^a }{d\tau}   \frac{d x^b }{d\tau}  \gamma_{ab}(x(\tau)) \,\,=\,\, \frac{1}{2} \int^{r_2}_{r_1} dr \, \gamma_{\perp\perp}(r) \q .
 \ea

 For a solution generated by a diffeomorphism  parametrized by a vector field $\xi^a$, the first order  metric perturbation is given by
 \ba
 \gamma_{\perp\perp} \,=\, ( {\cal L}_\xi  g )_{\perp\perp} \, =\,  \xi^a \p_a g_{\perp\perp} + 2 g_{\perp b} \p_\perp \xi^b \,\underset{(\ref{bgmetric}) }{=}\,  2 \p_\perp \xi^\perp.
 \ea
 We thus find
 \ba
 \ell &=& \xi^\perp(r_2) -\xi^\perp(r_1) \q .
 \ea
With  (\ref{xiofg}) and (\ref{2Ric}) we can express the $\xi^\perp$ component as a functional of the boundary metric
\ba
\xi^\perp\,=\, \frac{1}{\Delta} \Pi^{AB} \gamma_{AB} \,=\,  \frac{1}{\Delta } \delta ({}^2\! R)   \q .
\ea
But we see that the  geodesic lengths needs the metric $\gamma_{AB}$ at the outer boundary at $r_{\rm out}$ and at the inner boundary at $r_{\rm in}$.  In the following we will assume that $r_{\rm in}=0$ describes a one-- or zero--dimensional locus, that is a central axis or point. We will later see that in these cases, making certain smoothness assumptions on the metric perturbations and Fourier transforming in the spatial $y^A$ coordinates, $\xi^\perp(r=0)$ is indeed vanishing for almost all momentum modes. The following will hold for momentum modes for which $\xi^\perp(r=0)$ is vanishing. For these modes we have that $\ell=\xi^\perp(r_{\rm out})$ is a functional of the (outer) boundary metric only.
 
 Now consider the action 
 \ba \label{baction1}
\kappa  S_{\rho}\,=\, \frac{1}{4}  \int d^2 y \sqrt{h}  \left( \rho \,   \Delta_{} \rho  - 2 \rho \,   \delta ({}^2\! R) \right)   \q .
 \ea
Its equation of motion
 \ba
 \rho \,=\, \frac{1} \Delta  \delta ({}^2\! R) \,=\, \xi^\bot
 \ea
shows that on-shell $\rho=\ell$, and that the on-shell action
 \ba
\kappa  S_{\rho}\,\underset{\text{solu}}{=}\,   -\frac{1}{4}\int d^2 y \sqrt{h}  \,\xi^\bot \Delta_{} \xi^\bot
 \ea
 does indeed reproduce the boundary tangential invariant part of the gravitational Hamilton--Jacobi functional. 
 
 The action (\ref{baction1}) is local, with a quadratic term defined by $\Delta= 2 (K^{CD} - K h^{CD}) D_C D_D - {}^2\! R \,K$ and a Liouville-like coupling to the Ricci--scalar of  the boundary.
 
 In section \ref{BigSecEffGL} we will derive an effective action for the geodesic length observable more directly from the gravitational action. That is, we integrate out from the gravitational action all fields excepts for a degree of freedom describing the geodesic length. This resulting effective action will be very similar to (\ref{baction1}), but there will be also a non--local modification. This modification will take into account that $\xi^\perp(r=0)$ might be non--vanishing for certain momentum modes.

 \subsection{Action for the boundary tangential diffeomorphisms}\label{secTang}

So far we have found a boundary theory which reproduces the boundary diffeomorphism invariant part of the gravitational on-shell action. Its equation of motion for the field $\rho$ imposes that $\rho=\xi^\perp$, where $\xi^\perp$ is understood as a functional of the boundary metric. Similarly we can find an action which reproduces the remaining parts of the gravitational on-shell action, which are quadratic in the tangential boundary diffeomorphism parameters $\xi^A$. The dynamical variable is a boundary vector field $\sigma^A$ and the equations of motion will impose that $\sigma^A=\xi^A$.

To this end remember that the relation between $\xi^A$ and the boundary metric perturbations is given by
\ba
 {{\cal D}^A}_B \,  \xi^B &=&2(K^{BC}-K h^{BC}) \, \delta\, {}^2\! \Gamma^A_{BC} \q .
\ea
The action 
\begin{align}
 -\kappa  S_\sigma =   \frac14 \int d^2 y \sqrt{h} \left(  \sigma^A    {\cal D}_{AB}    \sigma^B - 4 \sigma^A h_{AD} (K^{BC}-K h^{BC})\, \delta\, {}^2\!\Gamma^D_{BC} \right)\\
\end{align}
leads to the equation of motion
\ba
 {\cal D}_{AB}    \sigma^B = 2  h_{AD} (K^{BC}-K h^{BC}) \delta \Gamma^D_{BC}
\ea
which are solved by $\sigma^A = \xi^A$.  On-shell the action evaluates to
\begin{align}
 -\kappa  S_{\sigma}\,\underset{\text{solu}}{=}\,  - \frac14     \int d^2 y \sqrt{h} \,   \xi^A {\cal D}_{AB} \xi^B.
\end{align}

Hence we can define a boundary theory, with three dynamical fields $\rho, \sigma^1,\sigma^2$, 
\begin{align}
 -\kappa S_{(\rho,\sigma)} =\tfrac{1}{4}  \int d^2 y \sqrt{h} \,  \left(-   \rho \,   \Delta_{} \rho   +   \sigma^A    {\cal D}_{AB}    \sigma^B                   + 2 \rho \,   \delta ({}^2\! R)- 4 \sigma_A (K^{BC}-K h^{BC}) \delta {}^2\!\Gamma^A_{BC} \right) 
\end{align}
which reproduces the second order gravitational on-shell action
 \ba
 -\kappa  S^{(2)}_{\rm HJ}&=& \frac{1}{4} \int d^2 y   \sqrt{h}  \, \left(\xi^\bot \Delta \, \xi^\bot \, -  \, \xi^A {\cal D}_{AB} \,\xi^B  \right) \q .
\ea

 \section{The effective action for the geodesic length}\label{BigSecEffGL}

We have seen that we can postulate an action for a boundary field theory, such that the boundary field variable evaluates to the geodesic lengths on solutions, and the action reproduces the (boundary diffeomorphism invariant part of the) Hamilton--Jacobi functional of gravity. Later we will however encounter examples for which the postulated action will differ in some subtle ways from the effective action for the geodesic lengths. This effective action is obtained by   integrating out all degrees of freedom from the gravitational action, except those parametrizing the geodesic lengths. These differences concern in particular the proper reflection of the (gauge) symmetries of the theory, and are, as we will discuss,  in particular important for the one-loop correction for the gravitational partition function. 
  
Integrating out all variables except for the geodesic lengths is hard to do directly\footnote{This can be achieved in Regge calculus \cite{BonzomDittrich,ADH}, but requires to employ a discretization of the theory, which might break the underlying diffeomorphism symmetry for backgrounds with curvature \cite{DittrichReview08,BahrDittrich09a}.}, as the geodesic lengths is a composite observable in terms of the metric perturbations. Instead we will add a Lagrange multiplier term to the second order action, 
 \ba\label{ActionLambda}
-\kappa  S^{(2)}_\lambda
&=& 
\frac{1}{4} \int_M d^3 x  \sqrt{g} \, \,\gamma_{a b} \left(   
  V^{abcd} \,
  \gamma_{cd} \,\,+ \,\,
  \tfrac{1}{2}  \,
  G^{abcdef }
   \, \nabla_c \nabla_d \gamma_{ef}  \right) \,+ \nn\\
   &&
  \frac{1}{4} \int_{\partial M} d^{2} y   \sqrt{h}\,\epsilon \,
\gamma_{ab}  \left(\,(B_1)^{abcd}\gamma_{cd} +   \, (B_2)^{abcde} \nabla_c \gamma_{de} \right)  \,+\nn\\
 && \frac{1}{4} \int_{(\partial M)_{\rm out}} d^{2} y\, \, \lambda(y) \left( \rho(y) - \ell[ \gamma_{\perp\perp}] \right)
\ea
 where $\lambda$ is a scalar density with respect to the boundary metric, which we treat as first order variable. The boundary field $\rho$ is a scalar, and the $\lambda$ equation of motion imposes that, evaluated on solutions, it gives the geodesic lengths
 \ba
 \ell= \frac{1}{2}\int^{r_{\rm out}}_{r_{\rm in}} dr \,\, \gamma_{\perp \perp} \q .
 \ea
 
 Here we allow for now to have either one outer boundary or one outer and an inner boundary. In the latter case we consider geodesics which go from the point $(r_{\rm out},y)$ on the outer boundary to the point $(r_{\rm in}, y)$ on the inner boundary. In the case where we have only an outer boundary the geodesic goes from $(r_{\rm out},y)$ to a bulk point $(r=0,P_{r\rightarrow 0}(y))$ where $P_{r\rightarrow 0}(y)$ is a projection of the $y$--coordinate to the set of points described by $r=0$. E.g. if we have a cylindrical set up with coordinates $(r,t,\theta)$ we have $P_{r\rightarrow 0}(t,\theta)=t$ describing a point along the axis $(r=0,t)$. 

Varying the action (\ref{ActionLambda}) with respect to the metric components we find the equations of motion
\ba\label{EOMlambda}
\hat G^{ab}:=\left(   
  V^{abcd} \,
  \gamma_{cd} \,\,+ \,\,
  \tfrac{1}{2}  \,
  G^{abcdef }
   \, \nabla_c \nabla_d \gamma_{ef}  \right)&=& \frac{1}{4} \frac{\lambda(y)}{\sqrt{h}} \delta^{a}_\perp \delta^b_\perp   \q ,
\ea
 where we have used that with our choice of Gaussian coordinates $\sqrt{g}=\sqrt{h}$.
 
At this point one might wonder about the fate of the contracted Bianchi identities
\ba
\nabla_a \hat G^{ab}=0
\ea
 which guarantee that three of the (vacuum) Einstein equations are redundant. But the divergence is also vanishing for the right hand side of (\ref{EOMlambda})
 \ba
 \nabla_a \frac{\lambda(y)}{\sqrt{h}} \delta^{a}_\perp \delta^b_\perp&=&\left( \lambda(y) \partial_\perp \frac{1}{\sqrt{h}}  +  \frac{\lambda(y)}{\sqrt{h}} \Gamma^A_{ A\perp} \right) \delta^\perp_b\,=\, 0 \q .
 \ea
Hence we still have three redundancies between the six equations of motion. We can therefore expect to be able to solve for the three metric components $\gamma_{\perp\perp}$ and $\gamma_{\perp A}$ in terms of the `spatial' metric $\gamma_{AB}$ and $\lambda$. In the examples, we will consider in the following, it is sufficient to consider the  three equations  (\ref{EOMlambda}) for $a=\perp$ and $b=\perp,A$.  Putting back the (possible $\lambda$--dependent) solutions for lapse and shift into $\hat G^{AB}=0$, one will find that these are automatically satisfied.
 
 In the following we will consider three examples:  a torus boundary embedded into flat space, a torus boundary embedded into hyperbolic (AdS) space, and a spherical boundary embedded into flat space. The cases with a torus boundary have  a boundary internal curvature ${}^2\! R=0$ and we will see that these cases are qualitatively different from the spherical boundary where ${}^2\! R\neq 0$. 

In particular for the cases with ${}^2\! R= 0$ the solution for the lapse $\gamma_{\perp\perp}$ resulting from (\ref{EOMlambda}) will not depend on $\lambda$. This applies in general for boundaries with ${}^2\! R= 0$ as will be shown in Appendix \ref{AppE}.  This prevents us from finding a solution for $\lambda$, and the resulting action will be simply the gravitational Hamilton--Jacobi functional with the Lagrange multiplier term added. 

There is however a resolution, if we consider only having an outer boundary and thus include $r=0$ into the bulk manifold $M$. In this case one has to take into account smoothness conditions for the metric perturbations at $r=0$. These conditions will constraint certain Taylor expansion coefficients of the `spatial' metric components $\gamma_{AB}$, and in case we have a Lagrange multiplier term,  render these $\lambda$--dependent. This mechanism will allow us to find an effective action for the geodesics lengths which can also serve as a gravitational dual boundary field theory.  The subtle point here is that certain properties of this boundary field theory  are determined by the smoothness conditions at $r=0$, even if we consider an asymptotic boundary $r_{\rm out}\rightarrow \infty$.

 \section{Twisted thermal  flat space with finite boundary}\label{SecFlat}
 
 As our first example we consider a background geometry known as twisted or spinning thermal flat space \cite{TwistedFlatSpace}. An effective action for the geodesic lengths has been found in \cite{BonzomDittrich} using a Regge discretization of gravity. This will allow us to compare and check the results obtained here.

 The metric of thermal spinning flat space is given by
\ba\label{metric0TSF}
ds^2&=& dr^2  + dt^2+ r^2 d\theta^2
\ea
with periodic identification $(r,t,\theta)\sim (r,t+\beta,\theta+\gamma)$ in addition to the usual identification $\theta\sim \theta+2\pi$ for the angular variable.

If we consider the spacetime for $0\leq r\leq r_{\rm out}$ we obtain a solid torus. Contractible cycles include curves described by $t=\text{const},\, r=\text{const}$ and non--contractible cycles include curves along  $\theta=\text{const}., \, r=\text{const}$. The torus can be obtained by identifying the top and bottom discs of a cylinder of height $\beta$, with a twisting angle (or angular potential) $\gamma$.   

The boundary extrinsic (background) curvature  is given by $K_{AB}=r\delta^\theta_A\delta^\theta_B$  and the boundary intrinsic (background) curvature is vanishing ${}^2\!R=0$.    Hence we have  a differential operator $\Delta=-2r^{-1} \partial_t^2$, which involves only derivatives in $t$--direction. 
 
 As the intrinsic curvature is vanishing we can define a Fourier transform for the metric perturbation components. We have to be however careful to implement the periodicity $(r,t,\theta)\sim (r,t+\beta,\theta+\gamma)$ of these functions into the Fourier transform. This can be done by `twisting' the phase factors for the Fourier transform so that these have the same periodicity:
 \ba\label{Ftrafo1}
 \gamma_{ab}(r,k_t,k_\theta)&=& \frac{1}{\sqrt{2\pi \beta}} \int^{\beta/2}_{-\beta/2}   dt  \int^{\pi}_{-\pi}   d\theta \, \gamma_{ab}(r,t,\theta) \,e^{-\im \theta k_\theta}e^{ - \im \frac{2\pi t}{\beta} ( k'_t-\frac{\gamma}{2\pi} k_\theta)} \q ,
 \ea
 where we will use the abbreviation $k_t:=\frac{2\pi }{\beta} ( k'_t-\frac{\gamma}{2\pi} k_\theta)$, and $k_\theta,k'_t \in {\mathbb Z}$. The inverse transform is given by
 \ba
 \gamma_{ab}(r,t,\theta)&=& \frac{1}{\sqrt{2\pi \beta}}  \sum_{k_t,k_\theta} \gamma_{ab}(r,k_t,k_\theta)  \,e^{\im \theta k_\theta}e^{  \im \frac{2\pi t}{\beta} ( k'_t-\frac{\gamma}{2\pi} k_\theta)}  \q .
 \ea
 
 \subsection{Equations of motion}
 
Using the Fourier transform the equations of motion (\ref{EOMlambda}) 
\ba\label{EOMExp}
\hat G^{ab}&=& \frac{1}{4} \frac{\lambda(y)}{\sqrt{h}} \delta^{a}_\perp \delta^b_\perp   \q ,
\ea
can be straightforwardly evaluated. The $(ab)=(\perp B)$ equations can be solved for the lapse and shift components $\gamma_{\perp\perp}$ and $\gamma_{\perp A}$ of the metric perturbations. (See also Appendix \ref{AppE}, which discusses the solutions for general backgrounds with flat spatial slices, that is with ${}^2R=0$.)

One finds 
\ba\label{LS1}
\gamma_{\perp\perp}\ &=& 2\partial_\perp \left( \frac{1}{2r} \left( \gamma_{\theta \theta} + \frac{k_\theta^2}{k_t^2} \gamma_{tt}-2 \frac{k_\theta}{k_t} \gamma_{\theta t}\right)\right) \nn\\
&=&2  \partial_\perp  \xi^\perp  \q ,\nn\\
\gamma_{\perp \theta} &=& \im k_\theta \frac{1}{2r} \left( \gamma_{\theta \theta} + \frac{k_\theta^2}{k_t^2} \gamma_{tt}-2 \frac{k_\theta}{k_t} \gamma_{\theta t}   \right) + r^2 \partial_\perp \left( \frac{\im}{r^2} \left( \frac{k_\theta}{2k_t^2} \gamma_{tt} - \frac{1}{k_t} \gamma_{\theta t}   \right) \right)  \,  -\im k_\theta  \lambda\frac{1}{4k_t^2}\nn \\
&=&\im k_\theta \xi^\perp + r^2 \partial_\perp \xi^\theta \, -\im k_\theta  \lambda\frac{1}{4k_t^2}\q , \nn\\
\gamma_{\perp t} &=& \im k_t\frac{1}{2r} \left( \gamma_{\theta \theta} + \frac{k_\theta^2}{k_t^2} \gamma_{tt}-2 \frac{k_\theta}{k_t} \gamma_{\theta t}\right) + \partial_\perp \left(-\frac{\im}{2k_t} \gamma_{tt} \right) \, - \im k_t \lambda\frac{1}{4k_t^2}\nn\\
&=&  \im k_t  \xi^\perp  + \partial_\perp \xi^t \, - \im k_t \lambda\frac{1}{4k_t^2} \q .
\ea
For $\lambda=0$ these confirm the relations (\ref{SolLapShif}) between the metric perturbations and the diffeomorphism generating vector $\xi^a$.  
Note also that the $\lambda$ dependence can be described by replacing $\xi^\perp$ by  
\ba\label{lambdashift0}
\hat \xi^\perp\, = \,\xi^\perp - \frac{1}{2\Delta} \frac{\lambda}{\sqrt{h}}\,=\,  \xi^\perp -\frac{1}{4k_t^2}\lambda  \q .
\ea

Using the solutions for lapse and shift perturbations in the remaining equations $\hat G^{AB}=0$, one finds that these are automatically satisfied, see also the discussion in section \ref{BigSecEffGL}.

Thus, if we are solving the equations for $r\in [r_{\rm in},r_{\rm out}]$ with $r_{\rm in}>0$ we can conclude that the `spatial' metric perturbations $\gamma_{AB}$ can be  freely chosen in the bulk. If we consider only an outer boundary and thus include $r=0$ in $M$, we will however argue that we have to impose some smoothness conditions on the metric components at $r=0$. We will see that this restricts certain Taylor expansion coefficients (arising from an expansion around $r=0$) of the spatial metric components. 

We have one remaining equation, coming from the variation of the Lagrange multiplier, namely
\ba\label{lambda0eq}
\rho&=& \frac{1}{2} \int_{r_1}^{r_2} dr \,  \gamma_{\perp\perp}\,=\, \xi^\perp(r_2)-\xi^\perp(r_1) \q ,
\ea
where\footnote{We could also write $\rho=\hat \xi^\perp(r_2)-\hat \xi^\perp(r_1)$ with a $\lambda$-dependent $\hat \xi^\perp(r)$ defined in (\ref{lambdashift0}). However, note that the $\lambda$-dependent terms drop out, as the $\lambda$--dependent term in $\hat \xi^\perp(r)$ is $r$-independent.}
\ba
\xi^\perp &=& \frac{1}{2r} \left( \gamma_{\theta \theta} + \frac{k_\theta^2}{k_t^2} \gamma_{tt}-2 \frac{k_\theta}{k_t} \gamma_{\theta t}\right)
\ea
does not depend on $\lambda$, at least not for non-vanishing radius.  
Considering the case with non-vanishing $r_1,r_2$ this equation only involves fixed boundary data and the field $\rho$, which we treat here as parameter, and not as a variable to solve for. There is no variable left, for which we can solve (\ref{lambda0eq}) and thus $\lambda$ remains a free parameter. 
 
 \subsection{Evaluating the action on solutions}\label{SecEvAc1}
 
We proceed by inserting the solutions (\ref{LS1}) into the action with Lagrange multiplier term (\ref{ActionLambda}). 
  Let us first consider the case that we have an outer boundary at $r_{\rm out}$ and an inner boundary at $r_{\rm in}$. 
 
  From the bulk term of the action we get a contribution 
  \ba
-\kappa  S^{(2)}_{\rm bulk}
\,=\, 
\frac{1}{4} \int_M d^3 x  \sqrt{g} \, \,\gamma_{a b} \  \,
 \hat  G^{ab} 
 &=& \frac{1}{16} \int_M \!\! d^2 y dr  \, \,\gamma_{\perp \perp} (r,y)  \lambda(y) 
 \,=\,\frac{1}{8} \int_{\partial M} \!\!\!\!  d^2y \, \lambda(y) \,\epsilon \xi^\perp  
   \; ,
\ea
where $\epsilon =+1$ for the outer boundary component and $\epsilon =-1$ for the inner boundary component.

The boundary terms split into two parts: firstly the part which arises from the vacuum solution (without $\lambda$), and secondly the part which appears due to the presence of $\lambda$. We have determined the first part $S^{(2)}_{\rm HJ}$ in (\ref{SHJ2}) (and Appendix \ref{App2HJ}. The $\lambda$--dependent part is derived in Appendix \ref{AppF}, where it is shown that it amounts also to a boundary integral over $\epsilon \lambda \xi^\perp$. We thus have
\ba
 -\kappa S^{(2)}_{\rm bdry}&=&  -\kappa  S^{(2)}_{\rm HJ}-\frac{1}{8} \int_{\partial M}  \!\!\!\! \!  d^2y \, \epsilon \lambda(y)  \xi^\perp \q .
\ea

 We see that the $\lambda$--dependent terms cancel from the gravitational action. We are left with the gravitational Hamilton--Jacobi functional and the Lagrange multiplier term
 \ba\label{EffA01}
 -\kappa  S^{(2)}_\lambda&\underset{\text{solu}}{=}&-\kappa  S^{(2)}_{\rm HJ} +
\,\frac{1}{4} \int_{(\partial M)_{\rm out}}  \!\!\!\! \! \!\!\!\! \! d^{2} y\, \, \lambda(y) \left( \rho(y) - \ell[ (\gamma_{AB})_{\rm out},(\gamma_{AB})_{\rm in} ] \right)\q \q
 \ea
 where the geodesic lengths $\ell$ is now understood as a functional of the boundary metric perturbations.
 
This is an effective action for the geodesic lengths as the boundary field $\rho$ evaluates to the geodesic length on solutions. But we cannot interpret (\ref{EffA01}) as a proper dual boundary field theory for gravity. 

Note that the same cancellation between the $\lambda$-dependent terms in the bulk and boundary contributions to the action seems to appear if we have only an outer boundary, that is if we consider as manifold $M$ the full solid torus. This however conflicts with the result of  \cite{BonzomDittrich}, which used a Regge calculus set-up. There the geodesic length variables can be explicitly identified with certain edge lengths, which serve as basic variables in Regge calculus. This allows to integrate out all variables except for those edge lengths identified with the geodesic lengths. This results in an effective action, which can be interpreted as a dual boundary field theory. 

In fact, adopting the approach of \cite{BonzomDittrich} to the case of an outer and inner boundary, that is to a torus ring, one finds the same result as in (\ref{EffA01}). As one now deals with a finite dimensional system one can identify the reason for this behaviour. To this end one splits the variables into two sets. The first set of variables ${\cal L}$ give the geodesic lengths, the other set ${\cal E}$ contains all remaining edge lengths. The linearized action has a Hessian with non--vanishing\footnote{Linearized Regge calculus on a flat background exhibits a remnant of the gauge symmetries of the continuum theory \cite{Williams,DittrichReview08, BahrDittrich09a}. But these gauge symmetries are associated to bulk vertices and one can triangulate the torus ring without any such bulk vertices, but nevertheless allow for an arbitrarily fine boundary triangulation. Thus one would not find gauge symmetries for this case. Note that the triangulation invariance of 3D linearized Regge calculus (and the associated one-loop partition function) \cite{steinhaus11} allows to use the coarsest possible bulk triangulation.} determinant, which allows to integrate out all variables. However, the subdeterminant associated to the variables ${\cal E}$ is actually vanishing. Thus we cannot integrate out straightforwardly all variables but the geodesic lengths. If one uses an action with a Lagrange multiplier term one will find that $\lambda$ remains a free parameter, and that the on-shell action is of the form (\ref{EffA01}), that is given by the Hamilton--Jacobi functional of the original system plus the Lagrange multiplier term. Appendix \ref{AppEff},  explains this general mechanism.

This opens the question why one does get a different result in Regge calculus for the case with just the outer boundary, that is for the solid torus \cite{BonzomDittrich}. The answer is, that in Regge calculus certain conditions, which guarantee the smoothness of the solution (in the continuum limit) around $r=0$ are automatically implemented. We will thus proceed by implementing similar smoothness conditions for the continuum theory.

\subsection{Implementing smoothness conditions for the metric at $r=0$}\label{secsmooth}

The smoothness conditions we are going to impose arise from assuming Taylor expandable metric perturbations around the origin in Cartesian coordinates. After transformation to cylindrical coordinates we can deduce a certain behaviour in the radial coordinate $r$:
\ba\label{Expa1}
\gamma_{\perp \theta} &=&  a^{(1)}_{r\theta} \,  r \,+\, a^{(2)}_{r\theta} r^2 \,+\, O(r^3) \,,\nn\\
\gamma_{\theta\theta} &=&  \q\q  \q  \,\, a^{(2)}_{\theta\theta} \, r^2 \,+\, O(r^3) \,  ,\nn\\
\gamma_{\theta t} &=& a^{(1)}_{\theta t} r\,+\, a^{(2)}_{\theta t} \, r^2 \,+\, O(r^3) \, ,
\ea
and all other metric perturbations starting with $r^{0}$ terms. For a detailed derivation we refer to Appendix \ref{AppSmooth}. 

We will impose these conditions for the metric perturbations, also for the case that we include the Lagrange multiplier term. The same behaviour can be deduced from Regge calculus, if one studies which conditions on the variables one needs to impose, in order to reach  the Regge action for the solid torus from the Regge action for the torus ring in the limit where the inner radius goes to zero. 

We will see that we now need to consider three  separate cases, namely $|k_\theta|\geq 2$, $k_\theta=\pm 1$ and $k_\theta=0$. We will start with the generic case $|k_\theta| \geq 2$.

\subsubsection{ For modes $|k_\theta| \geq 2$}

For the convenience of the reader we again display the solutions for the lapse and shift variables (\ref{LS1}):
\ba\label{LS2}
\gamma_{\perp\perp}\ &=& 2\partial_\perp \left( \frac{1}{2r} \left( \gamma_{\theta \theta} + \frac{k_\theta^2}{k_t^2} \gamma_{tt}-2 \frac{k_\theta}{k_t} \gamma_{\theta t}\right)\right) \, ,\nn\\
\gamma_{\perp \theta} &=& \im k_\theta \frac{1}{2r} \left( \gamma_{\theta \theta} + \frac{k_\theta^2}{k_t^2} \gamma_{tt}-2 \frac{k_\theta}{k_t} \gamma_{\theta t}   \right) + r^2 \partial_\perp \left( \frac{\im}{r^2} \left( \frac{k_\theta}{2k_t^2} \gamma_{tt} - \frac{1}{k_t} \gamma_{\theta t}   \right) \right)  \,  -\im k_\theta  \lambda\frac{1}{4k_t^2} \, ,\nn \\
%
\gamma_{\perp t} &=& \im k_t\frac{1}{2r} \left( \gamma_{\theta \theta} + \frac{k_\theta^2}{k_t^2} \gamma_{tt}-2 \frac{k_\theta}{k_t} \gamma_{\theta t}\right) + \partial_\perp \left(-\frac{\im}{2k_t} \gamma_{tt} \right) \, - \im k_t \lambda\frac{1}{4k_t^2} \q .\nn\\
\ea
We Taylor expand all metric perturbations in $r$ and arrive at equations for the expansion coefficients $a^{(n)}_{ab}$. Imposing the conditions that $a^{(n)}_{ab}=0$ for $n<0$ and that $a^{(0)}_{a\theta}=0$ as well as $a^{(1)}_{\theta\theta}=0$ we arrive at the conclusions:
\begin{itemize}
\item In order for $a_{rr}^{(-2)}$ to vanish, we need 
\ba\label{Cond1}
\frac{k_\theta^2}{k_t^2} a^{(0)}_{tt} =0 \q .
\ea
Thus, we have $a^{(0)}_{tt}=0$ for $k_\theta\neq 0$. This also ensures that $a_{r\theta}^{(-1)}$ and $a_{rt}^{(-1)}$ vanishes. 

\item Notice that, according to the first equation in (\ref{LS1}) the coefficient $a_{rr}^{(-1)}$ vanishes and we do allow for  non--vanishing $a_{rr}^{(0)}$. The remaining requirement comes from demanding that $a_{r\theta}^{(0)}$ is vanishing. This leads to the equation (for $k_\theta \neq 0$)
\ba\label{Cond2}
\left(1-\frac{1}{k_\theta^2}\right) \left( \frac{k_\theta^2}{k_t^2} a_{tt}^{(1)} -2  \frac{k_\theta}{k_t} a^{(1)}_{\theta t} \right)\,=\, \frac{\lambda}{2 k_t^2} \q .
\ea
\end{itemize}

In summary we obtain the conditions (\ref{Cond1}) and (\ref{Cond2}) for the boundary components of the metric. We also see that we need a special treatment for the case $k_\theta=0$ and $k_\theta =\pm 1$.  (The case $k_t^2=\tfrac{4\pi^2 }{\beta^2} ( k'_t-\tfrac{\gamma}{2\pi} k_\theta)^2 =0$, which arises for rational values for $\tfrac{\gamma}{2\pi}$ will be discussed in section \ref{Seckt0}.)

Note that both (\ref{Cond1}) and (\ref{Cond2}) are a restriction on expansion coefficients for the spatial metric perturbations.  These conditions also determine the value of the $r$--component $\xi^\perp$ of the diffeomorphism generating vector field at $r=0$,
\ba
\xi^\perp(0)&=&  \lim_{r\rightarrow 0} \frac{1}{2r} \left(  \gamma_{\theta \theta}(r) +\frac{k_\theta^2}{k_t^2} \gamma_{tt}(r)-2 \frac{k_\theta}{k_t} \gamma_{\theta t}(r)\right)            \nn\\
&=&\q\q
\frac{1}{2} \left( \frac{k_\theta^2}{k_t^2} a^{(1)}_{tt}-2 \frac{k_\theta}{k_t} a^{(1)}_{\theta t}\right) \nn\\
&=&\q\q  \frac{1}{4}\frac{k_\theta^2}{(k_\theta^2-1)} \frac{\lambda}{ k_t^2}  \q 
\ea
which now is $\lambda$--dependent. 

Thus,  considering the equation of motion imposed by the Lagrange multiplier, we now find
\ba\label{LagrEq1}
\rho&=& \frac{1}{2}\int^{r_{\rm out}}_0 dr \, \gamma_{\perp\perp}(r) \,=\,  \int^{r_{\rm out}}_0 dr\, \partial_\perp \xi^\perp (r) \,=\, \xi^\perp(r_{\rm out}) -\xi^\perp(0) \q ,
\ea
where 
\ba
\xi^\perp(r_{\rm out} )\,=\,  \frac{1}{2r_{\rm out}} \left( \gamma_{\theta \theta}(r_{\rm out}) + \frac{k_\theta^2}{k_t^2} \gamma_{tt}(r_{\rm out})-2 \frac{k_\theta}{k_t} \gamma_{\theta t}(r_{\rm out})\right)
\ea
is a function of the boundary data.  As $\xi^\perp(0)$ is now $\lambda$--dependent, we do obtain a solution for the Lagrange multiplier
\ba\label{sollambda1}
\lambda &=&   4 k_t^2   \left(1-\frac{1}{k_\theta^2}\right)    \left(\xi^\perp(r_{\rm out})-\rho\right)  \q .
\ea
Note that on-shell of the solutions to the effective action for $\rho$, we will have $\rho=\xi^\perp(r_{\rm out})$ and thus for $|k_\theta|\geq 2$  vanishing $\lambda$  as well as a vanishing component $\xi^\perp(0)$.

~\\ 

The evaluation of the action proceeds similarly as in section \ref{SecEvAc1}. The bulk term still leads to
\ba
-\kappa S^{(2)}_{\rm bulk}&=&\frac{1}{8} \int_{\partial M}  \!\!\!\! d^2y \,\lambda(y) \,(\xi^{r}(r_{\rm out},y) - \xi^\perp(0,y)) \q ,
\ea
where we have used $\epsilon =+1$ as we have only the outer boundary.
The boundary term  gives
\ba
 -\kappa S^{(2)}_{\rm bdry}&=&  -\kappa  S^{(2)}_{\rm HJ}(r_{\rm out})  -\frac{1}{8} \int_{\partial M}  \!\!\!\! \!  d^2y \, \lambda(y) \xi^{r}(r_{\rm out},y)  \q .
\ea
The Lagrange multiplier term vanishes on the solutions  to  (\ref{LagrEq1}).

Thus the terms with $\lambda\xi^{\perp}(r_{\rm out})$ still cancel, but we remain with the $\lambda\xi^\perp(0)$ term. We therefore obtain
\ba
 -\kappa  S^{(2)}_\lambda&\underset{\text{solu}}{=}&-\kappa  S^{(2)}_{\rm HJ}(r_{\rm out}) -\frac{1}{8} \int_{\partial M}\!\!\!\! d^2y\, \lambda(y)\xi^\perp(0,y)\nn\\
 &=&-\kappa  S^{(2)}_{\rm HJ}(r_{\rm out}) +\frac{1}{2} \int_{\partial M}\!\!\!\! d^2y    \left(\xi^\perp(r_{\rm out})-\rho\right)
  \partial_t^2   \left(1+\frac{1}{\partial_\theta^2}\right)    \left(\xi^\perp(r_{\rm out})-\rho\right) \q\q \nn\\
  &=&
  -\kappa  S^{(2)}_{\rm HJ}(r_{\rm out}) +\frac{1}{2} \int_{\partial M}\!\!\!\! d^2y \, \xi^\perp(r_{\rm out})  \partial_t^2   \left(1+\frac{1}{\partial_\theta^2}\right) \xi^\perp(r_{\rm out}) + \nn\\
&&\,\, \,\, \frac{1}{2} \int_{\partial M}\!\!\!\! d^2y \,\left( \rho \, \partial_t^2   \left(1+\frac{1}{\partial_\theta^2}\right) \rho -2 \rho \, \partial_t^2   \left(1+\frac{1}{\partial_\theta^2}\right)   \xi^\perp(r_{\rm out}) \right)\q .
\ea
The Hamilton-Jacobi functional is given by (remember that $\Delta=-2r^{-1} \partial_t^2$)
\ba
-\kappa  S^{(2)}_{\rm HJ}(r_{\rm out})&=& \frac{1}{4} \int_{\partial M}\!\!\!\! d^2y \sqrt{h} \left(\xi^\perp \Delta \xi^\perp  -  \, \xi^A {\cal D}_{AB} \,\xi^B  \right)
\,=\, -\frac{1}{2} \int_{\partial M}\!\!\!\! d^2y \left( \,\xi^\perp \partial_t^2 \xi^\perp -  \xi^A h_{AB} \partial_t^2 \xi^B \right)\q\q\nn\\
\ea
and with $\xi^\perp=\Delta^{-1} \delta( {}^2\!R)\,=\, -2^{-1} r \partial_t^{-2} \delta ({}^2\!R) $ we can write
\ba\label{FlatOnSh}
 -\kappa  S^{(2)}_\lambda&\underset{\text{solu}}{=}& -
  \frac{1}{4} \int_{\partial M}\!\!\!\! d^2y \sqrt{h} \,\left( \rho \, \Delta  \left(1+\frac{1}{\partial_\theta^2}\right) \rho -2 \rho \,    \left(1+\frac{1}{\partial_\theta^2}\right) \delta ({}^2\!R) \right)+ \nn\\
&&\q  \frac{1}{4} \int_{\partial M}\!\!\!\! d^2y \sqrt{h} \left(\xi^\perp \Delta \frac{1}{\partial^2_\theta} \xi^\perp  -  \, \xi^A {\cal D}_{AB} \,\xi^B  \right) \q .
\ea
This does define an action for the boundary field $\rho$, whose on-shell value does  reproduce the gravitational Hamilton--Jacobi function $S_{\rm HJ}$.

We note that
\ba\label{DualActA}
\kappa  S'_\rho&:=&   \frac{1}{4} \int_{\partial M}\!\!\!\! d^2y \sqrt{h} \,\left( \rho \, \Delta  \left(1+\frac{1}{\partial_\theta^2}\right) \rho -2 \rho \,    \left(1+\frac{1}{\partial_\theta^2}\right) \delta ({}^2\!R) \right)
\ea
differs from the action $S_\rho$ which we found in section \ref{secGeod} by the insertion of the non--local operator $(1+\partial_\theta^{-2})$. (It does also reproduce $S^{(2)}_{\rm HJ}$ multiplied with this factor.) This insertion has an important consequence: the effective action for the geodesic lengths $S'_\rho$ does vanish for modes $k_\theta=\pm 1$. The effective action is furthermore ill--defined for $k_\theta=0$.

As we will see shortly, the modes $k_\theta=\pm 1$ have a special status, as we can have in this case a non-vanishing $\xi^\perp(r=0)$  (for vanishing $\lambda$). It can be expressed as a function of the spatial metric components. But in the bulk these are gauge degrees of freedom. We therefore cannot determine the geodesic length at $k_\theta=\pm 1$ from the boundary data. In fact, in the Regge calculus set-up \cite{BonzomDittrich} the geodesic length variables at $k_\theta=\pm 1$ can be identified with gauge parameters resulting from the residual diffeomorphism symmetry of Regge calculus  \cite{Williams, DittrichReview08, BahrDittrich09a}. For this reason the effective action for the geodesic length should vanish --- and we show below that it in fact does.

For the case $k_\theta=0$ one has also a  diffeomorphism generating vector field, which does not need to vanish at $r=0$. But this time it is the component $\xi^t$ that does not need to vanish and can be furthermore identified as a gauge parameter. We will see that here we are back to a situation similar to what we described for the case with two boundaries: one cannot straightforwardly integrate out all variables except the geodesic lengths and the on-shell value of the action $S_\lambda$ will reproduce the Hamilton--Jacobi functional and the Lagrange multiplier term.

The special status of these modes is also reflected in the one--loop  partition function for gravity, which reproduces the vacuum character of the BMS group \cite{barnich1502, BonzomDittrich}. As we will discuss shortly in section \ref{seconeloop} the one--loop determinant for gravity does coincide with the one--loop determinant of the boundary field theory (\ref{DualActA}). Here it is important that this determinant does only include a product over the modes $k_\theta \geq 2$, as the modes $k_\theta=0$ and $k_\theta=\pm 1$ do describe gauge degrees of freedom \cite{BonzomDittrich}.

\subsubsection{ For modes with $k_\theta=0$}

For $k_\theta=0$ we obtain the following solutions for the lapse and shift components
\ba\label{LS10}
\gamma_{\perp\perp}\ &=& 2\partial_\perp \left( \frac{1}{2r} \gamma_{\theta \theta} \right)  
\;\;\,=\, 2  \partial_\perp  \xi^\perp \q ,\nn\\
\gamma_{\perp \theta} &=&  r^2 \partial_\perp \left( -\frac{\im}{r^2}\frac{1}{k_t} \gamma_{\theta t}  \right) 
\;\;\,=\, r^2 \partial_\perp \xi^\theta \q ,  \nn\\
\gamma_{\perp t} &=& \im k_t\frac{1}{2r} \gamma_{\theta \theta} + \partial_\perp \left(-\frac{\im}{2k_t} \gamma_{tt} \right) \, - \im k_t \lambda\frac{1}{4k_t^2}
\;\;\,=\,  \im k_t  \xi^\perp  + \partial_\perp \xi^t \, - \im k_t \lambda\frac{1}{4k_t^2} \q .
\ea
We see that for $k_\theta=0$ we can have $a_{tt}^{(0)}\neq 0$.   This is the only non--vanishing component of the spatial metric $\gamma_{AB}$ at $r=0$, and as it remains arbitrary, should be understood as gauge parameter. Note that this (additional) gauge parameter only appears for $k_\theta=0$, as it is forced to vanish for $k_\theta \neq 0$ by the equations of motion. 

We also see that the vector field component $\xi^\perp$ does vanish at $r=0$.  The requirement that $a^{(0)}_{r\theta}$ vanishes, imposes $a^{(1)}_{\theta t}=0$. From the last equation in (\ref{LS10}) we obtain that the only $\lambda$--dependent shift component is given by
\ba
a^{(0)}_{rt}\,=\, -\frac{\im}{2k_t} a^{(1)}_{tt} \, - \im k_t \lambda\frac{1}{4k_t^2} \q .
\ea

From the Lagrange multiplier equation we obtain
\ba\label{RhoEqu0}
\rho\,=\, \xi^\perp(r_{\rm out}) \,=\,  \frac{1}{2r_{\rm out}}  \gamma_{\theta \theta} (r_{\rm out}) \q ,
\ea
where different from the general case, we do not have a $\lambda$--dependent term as $\xi^\perp(r=0,k_\theta=0)=0$. 

Therefore  we cannot determine $\lambda$ as a function of  $\rho$ and the boundary variables.   We are now in the same situation as described for the case with two boundaries in section \ref{SecEvAc1}. 
One can compute explicitly that the evaluation of the action yields the same result as in this case, namely 
\ba\label{Akth=0}
 {\kappa  S^{(2)}_\lambda}_{|k_\theta=0}\,\, \,&\underset{\text{solu}}{=}&\,\,\,{\kappa  S^{(2)}_{\rm HJ}}_{|k_\theta=0}\,\,  +\,\,\tfrac{1}{4} {\lambda \left( \rho- \xi^\perp(r_{\rm out}) \right)}_{|k_\theta=0}  \q .
\ea

\subsubsection{ For modes with $k_\theta=\pm 1$}

Here we find from equation (\ref{Cond2}) that 
\ba\label{Cond20}
\left(1-\frac{1}{k_\theta^2}\right) \left( \frac{k_\theta^2}{k_t^2} a_{tt}^{(1)} -2  \frac{k_\theta}{k_t} a^{(1)}_{\theta t} \right)\,=\, \frac{\lambda}{2 k_t^2} \, \stackrel{!}{=}\,0 \q ,
\ea
and thus $\lambda=0$. 
The vector field component $\xi^\perp(0)$ does not need to vanish and is given by
\ba
\xi^\perp (0)  \,=\, \frac{1}{2}\left( \frac{k_\theta^2}{k_t^2} a_{tt}^{(1)} -2  \frac{k_\theta}{k_t} a^{(1)}_{\theta t} \right) \q .
\ea
Thus we have for the geodesic length variable
\ba
\rho&=& \xi^\perp(r_{\rm out})-\xi^\perp(0)   \q .
\ea
But here we should understand $\xi^\perp(0)$ as a bulk variable -- in fact it is a gauge parameter, that only appears at $k_\theta =\pm 1$. 

Inserting the solutions into the action, we will have due to $\lambda=0$, that 
\ba\label{Akth=1}
 \kappa  S^{(2)}_\lambda(k_\theta=0) &\underset{\text{solu}}{=}&\kappa  S^{(2)}_{\rm HJ}(k_\theta=0) 
\ea
and that thus the effective action for the boundary field $\rho$ vanishes. This is also confirmed in the Regge calculus setting \cite{BonzomDittrich}.

\subsection{Modes with $k_t=0$}\label{Seckt0}

Remember that we have defined $k_t:=\frac{2\pi }{\beta} ( k'_t-\frac{\gamma}{2\pi} k_\theta)$. Thus, if $\gamma$ is a rational multiple of $2\pi$ there will be certain $k'_t,k_\theta \in \mathbb{Z}$ for which $k_t=0$.  At these angles and for such modes with $k_t=0$ we do not have a well--posed boundary problem, that is solutions do not exist for all possible boundary metric fluctuations.\footnote{The boundary fluctuations for which one can and cannot find solutions for lapse and shift can be read off from (\ref{LS2}). Eg. allowing only for non--vanishing fluctuations $\gamma_{\theta\theta}$ still allows for a solution.} The condition of rational $\gamma$ can be translated in how geodesics along the torus would wind around this torus, see \cite{DGLR1,DGLR2}. 

These modes with $k_t=0$ will lead to divergencies of the one--loop correction, which appear for all rational angles. This can be treated with an ad-hoc regularization, as in \cite{barnich1502}. Alternatively, one can use a discretization, e.g. Regge calculus as in \cite{BonzomDittrich} or the Ponzano--Regge model as in \cite{DGLR1,DGLR2,DGLR3}. Such a discretization allows only rational angles $\gamma$, but the discretization does introduce a cut--off. For a given rational angle, there is a choice of (minimal) discretization, for which such modes with $k_t=0$ do not appear. 

As discussed in \cite{DGLR1,DGLR2} the appearance of such divergencies seems to be an artifact of the linearization, or in the quantum theory an artifact of the semi-classical (or one-loop) approximation, at least if one considers a boundary with finite radius. \cite{DGLR1} shows however that the exact partition function, for a particular choice of boundary conditions, does reproduce the divergence structure in  the limit to infinite radius.

\subsection{The limit of large radius}

We found as effective action for the geodesic length
\ba\label{DualActAb}
\kappa  S'_\rho&:=&   \frac{1}{4} \int_{\partial M}\!\!\!\! d^2y \sqrt{h} \,\left( \rho \, \Delta  \left(1+\frac{1}{\partial_\theta^2}\right) \rho -2 \rho \,    \left(1+\frac{1}{\partial_\theta^2}\right) \delta ({}^2\!R) \right)
\ea
which features a non-local operator $(1-1/\partial_\theta^2)$. If we fix however the physical wave lengths of the angular modes $r^{-2}\partial_\theta^2=\text{const.}=C$ we see that
\ba
\left(1+\frac{1}{r^2}\frac{1}{C}\right)\, \underset{r\rightarrow \infty}{\longrightarrow} 1\, ,
\ea
and the effective action becomes local and we recover the action $S_\rho$ proposed in section \ref{secGeod}. 

In this way we define (radial) scalings 
\ba
[k_\theta]=1 \,, \q  [\gamma_{t t}]=0  \, , \q [\gamma_{\theta \theta}]=2 \, , \q [\gamma_{\theta t}]=1 \q .
\ea
Then we have $[\xi^\perp]=1$ and $[\xi^\theta]=-1$ as well as $[\xi^t]=0$. We have also $[\Delta]=-1$ and $[{\cal D}_{\theta\theta}]=+1$ as well as $[{\cal D}_{tt}]=-1$. We thus find that $[\xi^\perp\Delta\xi^\perp]=+1$ comes with the dominant radial scaling, as compared to the terms which are not invariant under boundary tangential diffeomorphisms, which are given by $[\xi^\theta{\cal D}_{\theta\theta}\xi^\theta]=-1$ and $[\xi^t{\cal D}_{tt}\xi^t]=-1$. In this sense we have that for large radius the diffeomorphism invariant term $\xi^\perp \Delta \xi^\perp$ dominates.

\subsection{One--loop determinant of the dual boundary field theory}\label{seconeloop}

By construction we have that the dual action $S'_{\rho}$ reproduces the (boundary diffeomorphism invariant part of the) gravitational action -- modulo the insertion of $(1+\partial_\theta^{-2})$. (To compensate one adds the gravitational action with  $- \partial_\theta^{-2}$ inserted.) Here we will show that the dual action also reproduces the one--loop determinant of gravity, which has been computed in the continuum for asymptotic boundaries in \cite{barnich1502} and in the discrete for finite boundaries in \cite{BonzomDittrich}. 

To compute the one-loop determinant for $S'_{\rho}$ given in (\ref{DualActA}), we will adopt a simple lattice regularization  for the Hessian of the action, which is given by $k_t^2 (1-k_\theta^{-2})$:
\ba
k_\theta^2 &\rightarrow& \left(2-2\cos\left( \frac{2\pi}{N_\theta}\right) \right)^{-1}  \left(2-2\cos\left( \frac{2\pi}{N_\theta}\kappa_\theta\right) \right) \nn\\
k_t^2 &\rightarrow& \frac{N_t^2}{\beta^2} \left(2-2\cos\left( \frac{2\pi}{N_t} ( \kappa_t-\frac{\gamma}{2\pi}\kappa_\theta)\right) \right) \q ,
\ea
where $\kappa_\theta=0, \ldots ,N_\theta-1$ and $\kappa_t=0,\ldots N_t-1$. With this choice we still have that $(1-k_\theta^{-2})=0$ for $\kappa_\theta=\pm 1$. Now, as our dual action is only defined for $|k_\theta|\geq 2$ we consider
\ba
\prod_{\kappa_\theta=2}^{N_\theta-2} \left( 1 -\frac{2-2\cos\left( \frac{2\pi}{N_\theta}\right)  }{2-2\cos\left( \frac{2\pi}{N_\theta}\kappa_\theta\right) } \right)&=&\frac{1}{2+2\cos\left( \frac{2\pi}{N_\theta}\right) }
\ea
and 
\ba
\prod_{\kappa_y=0}^{N_t-1} \left(2-2\cos\left( \frac{2\pi}{N_t} ( \kappa_t-\frac{\gamma}{2\pi}\kappa_\theta)\right) \right)&=&2-2\cos(\gamma\kappa_\theta) \q .
\ea
Ignoring some inessential constants we therefore have
\ba
\prod_{\kappa_\theta=2}^{N_\theta-2}\prod_{\kappa_y=0}^{N_t-1} \frac{1}{ \sqrt{k_t^2 (1-k_\theta^{-2}) }} &\sim&
 \prod_{\kappa_\theta=2}^{N_\theta/2-1}\frac{1}{| 1-q^{\kappa_\theta}|^2}
\ea
where $q=\exp(i\gamma)$. This reproduces the one--loop determinant of the gravitational theory \cite{barnich1502,BonzomDittrich}. Thus the (only) essential contribution to the one--loop determinant arises from the degrees of freedom describing the geodesic lengths from the boundary to some central point. This confirms the interpretation of the action $S'_\rho$ as dual action for gravity.

Note that to get the correct result, it is essential to not to include the modes $k_\theta =0$ and $k_\theta =\pm 1$, which in our case follows from the appearance of the non-local operator $(1-k_\theta^{-1})$.  The exclusion of the  $k_\theta=\pm 1$ modes is a feature of the vacuum BMS character \cite{Oblak}, which is reproduced by the one-loop partition function for asymptotic boundaries \cite{barnich1502}.  Inserting a point particle in the centre, one rather expects a massive character. Indeed the insertion of a  point particle will break the diffeomorphism symmetry described by the $k_\theta=\pm 1$ modes. This can be also expected to happen in the current framework, as we would have to modify the smoothness conditions, which we introduced in section \ref{secsmooth}, and which were essential for obtaining a suitable dual action.

 \section{Twisted thermal AdS space with finite boundary}\label{SecAdS}

Next we will consider as background AdS space with metric
\begin{align}
	ds^2 = \d r^2 + \sinh^2 \!r \,d \theta^2+ \cosh^2 \!r \, d t^2  \q ,
\end{align}
where we have fixed $\Lambda=-1$. As for the flat space metric we impose the periodicity conditions $(r,t,\theta)\sim (r,t+\beta,\theta+\gamma)$ and $\theta\sim \theta+2\pi$ for the angular variable. This defines twisted thermal AdS space. The one-loop partition function for this background with asymptotic boundary has been computed from the gravity side in \cite{Giombi} and reproduces the vacuum character of the asymptotic symmetries of AdS$_3$ space \cite{BrownHenneaux}. This example has been intensively discussed in the literature, e.g. \cite{Henneaux, Maloney, CotlerJensen} and references therein. The derivation of the (Liouville) dual boundary field theory starts often with the Chern--Simons formulation of 3D gravity. One exception is \cite{CarlipBreakingDiffeoAdS}, which derives a dual boundary theory from the breaking of diffeomorphism symmetry at the asymptotic boundary. In fact the field introduced in \cite{CarlipBreakingDiffeoAdS} agrees (in the linearized theory) with the geodesic distance employed here. Our derivation of the dual field theory is  somewhat more direct and also applicable to finite boundaries. 

We will again consider a torus boundary at $r=r_{\rm out}$ and thus the  background intrinsic curvature of the boundary (which we constrained to be homogeneous) has to vanish ${}^2\! R=0$.

The computation of the effective geodesic action is very similar to the flat case, and we will therefore be brief. One again finds that one needs to invoke smoothness conditions at $r=0$ in order to obtain an effective action, which can also serve as dual boundary field theory. The modes $k_\theta=0$ and $k_\theta=\pm 1$ will also play a special role. 

One difference with the flat case is that the extrinsic curvature has now full rank 
\ba
K_{\theta\theta}=K_{tt}=\cosh r \sinh r \q , \q\q K_{\theta t}=0 \q  \text{and} \q K=\tanh r+ \coth r \q .
\ea
Thus 
\ba
\Delta = 2(K^{CD}-Kh^{CD})D_C D_D \,=\, \frac{-2}{\cosh\! r \sinh\! r} \left( \partial_\theta^2 + \partial_t^2\right)\,=\, \frac{-2}{\sqrt{h}}\left( \partial_\theta^2 + \partial_t^2\right)
\ea
is now non-degenerate. 

The Fourier transformation for the $y=(\theta,t)$ variables can be defined as for the flat background, see (\ref{Ftrafo1}), which allows us to invert the various differential operators.

 \subsection{Equations of motion and evaluation of the action}
 
 The equations of motion
 \ba\label{EOMExpAdS}
\hat G^{ab}&=& \frac{1}{4} \frac{\lambda(y)}{\sqrt{h}} \delta^{a}_\perp \delta^b_\perp   \q ,
\ea
resulting from varying $\gamma_{ab}$ of the Lagrange multiplier action (\ref{ActionLambda}) can be solved for the lapse and shift metric perturbations. The solutions are given by
\begin{align}\label{AdSsol}
	\gamma_{\perp\perp} 
	=& 2 \p_\perp \Big( \frac{ k_\theta^2 \gamma_{tt} + k_t^2 \gamma_{\theta\theta} - 2 k_t k_\theta \gamma_{t\theta} }{2\cosh\! r \sinh \!r (k_t^2 + k_\theta^2) } \Big)\\
	=& 2 \p_\perp \xi^\perp \q , \\
	\gamma_{\perp t} 
	=& \im k_t \frac{ k_\theta^2 \gamma_{tt} + k_t^2 \gamma_{\theta\theta} - 2 k_t k_\theta \gamma_{t\theta} }{2\cosh\! r \sinh\! r (k_t^2 + k_\theta^2) } + \cosh^2 \!r \p_\perp \Big( \frac{ - \im k_t \gamma_{tt} + \im k_t \gamma_{\theta\theta} - 2 \im k_\theta \gamma_{t\theta} }{2 \cosh^2\! r(k_t^2 + k_\theta^2)} \Big) - \frac{\im k_t \lambda}{4 (k_t^2 + k_\theta^2)}\\
	=& \im k_t \xi^\perp + \cosh^2 \!r \p_\perp \xi^t - \frac{\im k_t \lambda}{4 (k_t^2 + k_\theta^2)} \q , \\
	\gamma_{\perp \theta}
	=& \im k_\theta \frac{ k_\theta^2 \gamma_{tt} + k_t^2 \gamma_{\theta\theta} - 2 k_t k_\theta \gamma_{t\theta} }{2\cosh\! r \sinh\! r (k_t^2 + k_\theta^2) } + \sinh^2 \!r \p_\perp \Big( \frac{ \im k_\theta \gamma_{tt} - \im k_\theta \gamma_{\theta\theta} - 2 \im k_t \gamma_{t\theta} }{2 \sinh^2\! r(k_t^2 + k_\theta^2)} \Big) - \frac{\im k_\theta \lambda}{4(k_t^2 + k_\theta^2)}\\
	 =& \im k_\theta \xi^\perp + \sinh^2\! r \p_\perp \xi^\theta - \frac{\im k_\theta \lambda}{4(k_t^2 + k_\theta^2)}\q . 
\end{align}
Thus the lapse and shift perturbations arise by replacing $\xi^\perp$ with
\ba
\hat \xi^\perp\,=\, \xi^\perp- \frac{1}{2\Delta} \frac{\lambda}{\sqrt{h}} \,=\,\xi^\perp - \frac{1}{4(k_t^2 + k_\theta^2)} \q .
\ea
As for the flat case we have that the solution for $\gamma_{\perp\perp}$ a priori does not involve $\lambda$. Appendix \ref{AppE} shows that this will be always the case for foliations for which ${}^2\! R=0$. Thus we will also find here that for the case of an outer and inner boundary, $\lambda$ remains a free parameter and the action (\ref{ActionLambda}) evaluated on the solutions  (\ref{AdSsol}) will just reproduce the gravitational Hamilton--Jacobi functional plus the Lagrange multiplier term.

If we consider only the case of an outer boundary we have to impose smoothness conditions for $r=0$. Adopting the same strategy as for the flat case we choose to impose 
\begin{align}
	\gamma_{\perp \theta} =& r a^{(1)}_{r\theta} + r^2 a^{(2)}_{r\theta} + O(r^3)\, ,\\
	\gamma_{\theta\theta} =& r^2 a^{(2)}_{\theta\theta} + O(r^3)\, ,\\
	\gamma_{t\theta} =& r a^{(1)}_{t\theta} + r^2 a^{(2)}_{r\theta} + O(r^3) \q ,	
\end{align}
with the remaining  metric components starting with $a^{(0)}_{ab} r^0$ coefficients. 

Ensuring that $a^{(-2)}_{rr}=0$ requires again $k_\theta a_{tt}^{(0)}=0$. To make $a_{r\theta}^{(1)}$ vanish we need
\begin{align}
	\lambda = (k_\theta^2 - 1) ( 2 a_{tt}^{(1)} - 4 \frac{k_t}{k_\theta} a_{t\theta}^{(1)}) \q .
\end{align}

This leads to a non-vanishing vector component $\xi^\perp$ at $r=0$:
\begin{align}
	\xi^\perp(r\!=\!0) \,\, =\,\, \frac{1}{4} \frac{k_\theta^2}{(k_\theta^2 - 1)} \frac{\lambda}{ (k_t^2 + k_\theta^2)}, 
\end{align}
which allows us to solve the Lagrange multiplier equation  $\rho=\xi^\perp(r_{\rm out}) - \xi^\perp(r\!=\!0)$ for $\lambda$:
\begin{align}
\lambda = 4 (k_t^2 + k_\theta^2) \left(1 - \frac{1}{k_\theta^2}\right) (\xi^\perp(r_{\rm out}) - \rho) \q .
\end{align}

The evaluation of the action proceeds completely parallel to the flat case and we arrive at 
\ba
 -\kappa  S^{(2)}_\lambda&\underset{\text{solu}}{=}& -
  \frac{1}{4} \int_{\partial M}\!\!\!\! d^2y \sqrt{h} \,\left( \rho \, \Delta  \left(1+\frac{1}{\partial_\theta^2}\right) \rho -2 \rho \,    \left(1+\frac{1}{\partial_\theta^2}\right) \delta ({}^2\!R) \right)+ \nn\\
&&\q  \frac{1}{4} \int_{\partial M}\!\!\!\! d^2y \sqrt{h} \left(\xi^\perp \Delta \frac{1}{\partial^2_\theta} \xi^\perp  -  \, \xi^A {\cal D}_{AB} \,\xi^B  \right) \q .
\ea
where now $\Delta=\frac{-2}{\cosh\! r \sinh \! r} \left( \partial_\theta^2 + \partial_t^2\right)$ and $\sqrt{h}= \cosh \! r \sinh\! r$.

The cases $k_\theta=\pm 1$ and $k_\theta=0$ require again special attention. For $k_\theta=\pm 1$ we find that $\lambda=0$ and that thus the action for the field $\rho$ vanishes. For $k_\theta=0$ we have that $\xi^\perp(r\!=\!0)$ vanishes, and that thus $\lambda$ remains undetermined. The on-shell evaluation of the $\lambda$--action will therefore give the same result (\ref{Akth=0}) as in the flat case.

In summary we find that the action for the boundary field $\rho$ features the same insertion of the non-local differential operator $(1+\partial_\theta^{-2})$ as in the flat case
\ba\label{SRAdS}
 \kappa  S'_\rho&:=&
  \frac{1}{4} \int_{\partial M}\!\!\!\! d^2y \sqrt{h} \,\left( \rho \, \Delta  \left(1+\frac{1}{\partial_\theta^2}\right) \rho -2 \rho \,    \left(1+\frac{1}{\partial_\theta^2}\right) \delta ({}^2\!R) \right) \q .
\ea

\subsection{One loop correction from the dual field}\label{seconeloopAdS}

We have thus found an effective boundary action for the AdS background. The kinetic part is describing a free scalar field on a torus. Additionally we have the operator $(1+\partial^{-2}_\theta)$ but we have seen in section \ref{seconeloop}, that, apart from suppressing the $k_\theta=\pm 1$ modes, this operator does only contribute a constant to the one-loop partition function.  But the Laplace operator $\Delta\sim \partial_t^2 + \partial_\theta^2$  defined on the torus leads to the one-loop correction
\ba
\prod_{\kappa_\theta>2} \frac{1}{|1-q^{\kappa_\theta}|^2}
\ea
where $q=\exp(i\tau)$ with the torus modular parameter $\tau=\tfrac{1}{2\pi}(\gamma-i\beta)$. This agrees with the one-loop correction computed directly from gravity \cite{Giombi}.

\section{Flat space with spherical boundary}\label{SecSph}

We have seen that for the cases with flat boundaries, that is with ${}^2\! R=0$, we need to carefully take into account smoothness conditions at $r=0$, to obtain an effective action, which can also be interpreted as dual field theory. This effective action does however differ by the insertion of a non-local operator from the action, which we postulated in section \ref{secGeod}. This non-local operator plays an important role in transferring correctly the symmetries of the gravitational theory to the dual field theory. 

In appendix \ref{AppE} we show that for all cases with ${}^2\! R=0$, the solution for the lapse fluctuation, and therefore for the geodesic lengths, will {\it not} depend on the Lagrange multiplier $\lambda$. We can therefore expect that the mechanism for constructing the effective action is similar to the cases discussed here. That is we have to carefully consider smoothness conditions at $r=0$, and might have to expect the insertion of a non-local operator. 

Let us now consider a case with non-vanishing background intrinsic curvature ${}^2\! R\neq 0$. As we consider only boundaries with homogeneous curvature, we have to change the topology. We will choose a spherical one. Using Regge lengths one can argue that for a sphere  boundary the effective action for the geodesic length should be local\footnote{The reason is that the one-loop partition function for 3D  Regge calculus is bulk triangulation independent \cite{steinhaus11}. One can therefore choose the coarsest bulk triangulation available. For the spherical boundary one can choose a triangulation with only one bulk vertex and where all bulk edges go from  the boundary to  this bulk vertex. The edge lengths can therefore be interpreted as geodesic lengths and the Regge action, which is local, can be identified with the effective action for the geodesic lengths \cite{ADH}.} and that we thus might confirm the action we postulated in section \ref{secGeod}.

We choose as background metric
\ba\label{sphMetric}
ds^2\,=\, dr^2 + r^2 d\theta^2 + r^2 \sin^2\!\theta d\varphi^2
\ea
with spherical boundary  defined by $r=\text{const}$. The intrinsic boundary curvature is now non--vanishing ${}^2\!R=\tfrac{2}{r^2}$. We will see that this alters the computations in several ways from the cases with intrinsically flat boundary.

We have furthermore $K_{CD}=\tfrac{1}{2} K h_{CD}$ and thus $K^{AB}-Kh^{AB}=-\tfrac{1}{2} K$. This gives
\ba\label{Dsph}
\Delta &=& -\left(  K D^C D_C +  {}^2\! R K \right)\,\, \;\q\q\,=\,   -\tfrac{2}{r} \left( D^CD_C +\tfrac{2}{r^2}\right)\q , \nn\\
{\cal D}_{AB} &=& -\left( K D^C D_C  + \tfrac{1}{2} {}^2\! R K\right)  h_{AB}\,=\,  -\tfrac{2}{r} \left( D^CD_C +\tfrac{1}{r^2}\right)h_{AB}
\ea
with $K=\tfrac{2}{r}$ and ${}^2\!R=\tfrac{2}{r^2}$. 

As we have now intrinsic curvature, the differential operators $D_\theta$ and $D_\varphi$ are non--commuting and we cannot simultaneously diagonalize these operators. However one can use scalar, vector and tensor spherical harmonics, which allow for the diagonalization of the Laplacian $\square_{\rm Lap}=h^{AB}D_A D_B$ acting on scalars, vectors and second rank tensors.  Furthermore one has certain properties for the divergence of the vector and tensor harmonics as well as for the trace of the tensor harmonics, see appendix \ref{AppHarm}. 

We will however not need these harmonics for most of the discussion. It will be sufficient to know that we can find the inverse of the operators $\Delta$ and ${\cal D}_{AB}$, e.g. by using the spherical harmonics to diagonalize these operators.

\subsection{Solutions to the equations of motion}

We again start by solving the lapse and shift components of the equations of motion
 \ba\label{EOMExpSph}
\hat G^{ab}&=& \frac{1}{4} \frac{\lambda(y)}{\sqrt{h}} \delta^{a}_\perp \delta^b_\perp   \q ,
\ea
for the lapse and shift components of the metric perturbations. The derivation of the solutions is now more involved, due to the non-commutativity of the differential operators. We have collected the essential details in appendix \ref{AppSph} and reproduce here just the resulting solutions for lapse and shift:
\ba\label{E1110c}
 \gamma_{\perp\perp} &=& 2 \partial_\perp \Delta^{-1}  \left(  \Pi^{AB} \gamma_{AB} \right) - r^{-1}\Delta^{-1}\frac{\lambda}{\sqrt{h}}  \q   \nn\\
 &=&2 \partial_\perp  \left(\xi^\perp  -\frac{1}{2} \frac{1}{\Delta} \frac{\lambda} {\sqrt{h}} \right) \q , \nn\\
 \gamma_{\perp B} 
&=&   D_B \Delta^{-1} \left( \Pi^{CD} \gamma_{CD}\right) -\frac{1}{2}    D_B \Delta^{-1}\frac{\lambda}{\sqrt{h}}
+ h_{BA}\, \partial_\perp   \left( {\cal D}^{-1} \left(  - \frac{2}{r}h^{CD} \delta {}^2\!\Gamma^\circ_{CD}\right) \right)^A\nn\\
&=& D_B \left( \xi^\perp  -\frac{1}{2} \frac{1}{\Delta} \frac{\lambda} {\sqrt{h}} \right)  + h_{BA} \partial_\perp \xi^A \q .
\ea

We again find that the introduction of the Lagrange multiplier amounts to shifting the vector component $\xi^\perp$ to 
\ba
\hat \xi^\perp \,\, = \xi^\perp -\frac{1}{2} \frac{1}{\Delta} \frac{\lambda} {\sqrt{h}} \q .
\ea
But different from the cases with flat boundary we now have a $\lambda$--dependence for the lapse components $\gamma_{\perp\perp}$. Here it arises due to the fact that $\sqrt{h} \Delta$ is now $r$--dependent.

Let us also shortly discuss the smoothness conditions for the metric perturbations at $r=0$. Assuming  Taylor expandable metric perturbations in Cartesian coordinates and transforming these to spherical coordinates, see Appendix \ref{AppSmooth}, we find that $\gamma_{\perp\perp}$  has an expansion in the $r$--coordinate that starts with $r^0$, $\gamma_{\perp A}$ components start with an $r^1$--term and the $\gamma_{AB}$--components start with $r^2$. 

Now assuming that the $\gamma_{AB}$ components start with $r^2$ one will find that the solutions (\ref{E1110}) ensure that the remaining conditions are satisfied. This holds also if we do include a non--vanishing $\lambda$. To see this, one can use the scaling properties of the differential operators in $r$, e.g. 
\ba
\Delta=r^{-3} \tilde \Delta \q ,\q\q  {{\cal D}^A}_B=  r^{-3} {\tilde {\cal D}^A\,}_B \q ,\q\q \Pi^{AB}= r^{-4} \tilde \Pi^{AB}
\ea
where $\tilde{\cal O}$ is the operator ${\cal O}$ evaluated at $r=1$. 

Using these scaling properties we can also deduce that the vector component $\xi^\perp$ is vanishing at $r=0$, that is we have $\xi^\perp(r=0)=0$ as well as $\hat \xi^\perp(r=0)=0$.

Finally we  consider the Lagrange multiplier equation, which is now given by
\ba
\rho\,=\,   \frac{1}{2} \int^{r_{\rm out}}_{r_{\rm in}}\!\! \!\!\! dr \, \gamma_{\perp\perp} &=&  \hat \xi^\perp(r_{\rm out}) - \hat \xi^\perp(r_{\rm in}) \nn\\ 
&=&   \xi^\perp(r_{\rm out}) - \xi^\perp(r_{\rm in}) -\frac{ (r_{\rm out} -r_{\rm in}) }{ 2 \sqrt{\tilde h} \tilde \Delta } \lambda
\ea
Thus we obtain as a solution for $\lambda$
\ba\label{lambdaS}
\lambda&=& \frac{2 \sqrt{\tilde h} \tilde \Delta }{ (r_{\rm out} -r_{\rm in})}     \left( \xi^\perp(r_{\rm out}) - \xi^\perp(r_{\rm in}) -\rho \right) \q ,
\ea
where $\xi^\perp(r=0)=0$.

\subsection{Evaluation of the action}

Let us consider the case that we have an outer boundary at $r_{\rm out}$ and an inner boundary at $r_{\rm in}$. As we have $\hat \xi^\perp(\!r=0\!)=\xi^\perp(\!r=0\!)=0$, it will be straightforward to derive from this the case with only an outer boundary.

For the evaluation of the boundary we need to consider the bulk and boundary term in (\ref{ActionLambda}) -- the Lagrange multiplier term vanishes on solutions of (\ref{lambdaS}). We will however treat for the moment $\lambda$ as a variable, and only use the explicit solution for $\lambda$ at the very end.

The bulk term gives evaluated on solutions of (\ref{EOMExpSph}) 
  \ba
-\kappa  S^{(2)}_{\rm bulk}
&=& 
\frac{1}{4} \int_M d^3 x  \sqrt{g} \, \,\gamma_{a b} \  \,
 \hat  G^{ab} 
 \,=\, \frac{1}{16} \int_M \!\! d^2 y dr  \, \,\gamma_{\perp \perp} (r,y)  \lambda(y) \nn\\
 &=&\frac{1}{8} \int_{(\partial M)_{\rm out}} \!\!\!\! \!\!\!\! d^2y \, \lambda \,(\hat\xi^{\perp}(r_{\rm out}) - \hat\xi^\perp(r_{\rm in})) \q . 
\ea
Note that we now have $\hat \xi^\perp$ appearing, instead of just $\xi^\perp$. (In the cases with flat boundaries $(\hat \xi^\perp-\xi^\perp)$ is constant in $r$ and we could thus use $\xi^\perp$.)

For the boundary term we find (see Appendix \ref{AppF})
\ba
 -\kappa S^{(2)}_{\rm bdry}&=&  -\kappa S^{(2)}_{\rm HJ} -\frac{1}{8} \int_{(\partial M)_{\rm out}}  \!\!\!\! \! \!\!\! d^2y \, \lambda \left(\xi^\perp(r_{\rm out}) -\xi^\perp(r_{\rm in})\right)  \q .
\ea

We are thus left with 
 \ba\label{EffS1}
 -\kappa  S^{(2)}_\lambda&\underset{\text{solu}}{=}&-\kappa S^{(2)}_{\rm HJ}+ 
\frac{1}{8} \int_{(\partial M)_{\rm out}} \!\!\!\!\!\!\!\! d^{2} y\, \, \lambda \left( ( \hat \xi^\perp-\xi^\perp)(r_{\rm out}) - ( \hat \xi^\perp-\xi^\perp)(r_{\rm in}) \right)\nn\\
&=&-\kappa S^{(2)}_{\rm HJ}
 -\frac{1}{8} \int_{(\partial M)_{\rm out}} \!\!\!\!\!\!\!\!d^{2} y\, \, \lambda  \,\frac{(r_{\rm out}-r_{\rm in})}{2 \sqrt{\tilde h} \tilde \Delta} \, \lambda \q .
 \ea
Inserting the solution (\ref{lambdaS}) for $\lambda$
\ba\label{lambdaS1}
\lambda&=& \frac{2 \sqrt{\tilde h} \tilde \Delta }{ (r_{\rm out} -r_{\rm in})}     \left( \xi^\perp(r_{\rm out}) - \xi^\perp(r_{\rm in}) -\rho \right) \q ,
\ea
we obtain
 \ba\label{EffS2}
 -\kappa  S^{(2)}_\lambda&\underset{\text{solu}}{=}&-\kappa  S^{(2)}_{\rm HJ} 
 -\frac{1}{4} \int_{(\partial M)_{\rm out}} \!\!\!\!\!d^{2} y \frac{\sqrt{\tilde h}}{  (r_{\rm out} -r_{\rm in})   }
 \bigg[
 \rho \tilde \Delta \rho\,- 2\rho \tilde \Delta   \left( \xi^\perp(r_{\rm out}) - \xi^\perp(r_{\rm in}) \right) + \nn\\
 &&\q\q\q\q\q\q\q\q \q\q\q \left( \xi^\perp(r_{\rm out}) - \xi^\perp(r_{\rm in})  \right) \tilde \Delta   \left( \xi^\perp(r_{\rm out}) - \xi^\perp(r_{\rm in})  \right)
 \bigg] \,.\q\q
\ea
The terms in  $S^{(2)}_{\rm HJ}$, in which $\xi^\perp$ appears are given by
\ba
\sqrt{h} \xi^\perp(r_{\rm out}) \Delta \xi^\perp(r_{\rm out})&=& r^{-1}_{\rm out}\sqrt{\tilde h} \xi^\perp(r_{\rm out}) \tilde\Delta \xi^\perp(r_{\rm out}) \q \text{and} \nn\\ 
-\sqrt{h} \xi^\perp(r_{\rm in}) \Delta \xi^\perp(r_{\rm in})&=&- r^{-1}_{\rm in}\sqrt{\tilde h} \xi^\perp(r_{\rm in}) \tilde\Delta \xi^\perp(r_{\rm in})
\ea
Thus, for $r_{\rm in}\neq 0$ we will not have a cancellation between these terms and 
\ba
 (r_{\rm out} -r_{\rm in}) ^{-1}\sqrt{\tilde h}    \left( \xi^\perp(r_{\rm out}) - \xi^\perp(r_{\rm in})  \right) \tilde \Delta   \left( \xi^\perp(r_{\rm out}) - \xi^\perp(r_{\rm in})  \right)
\ea
appearing in (\ref{EffS2}).

Thus, although (\ref{EffS2}) is an effective action for the geodesic lengths between the outer and inner boundary, we cannot interpret the $\rho$--dependent part as a dual action for gravity.  This might not be a surprise as the geodesic lengths does only detect the difference between $\xi^\perp(r_{\rm out})$ and $\xi^\perp(r_{\rm in})$, whereas for the evaluation of the gravitational boundary term we need to know both $\xi^\perp(r_{\rm out})$ and $\xi^\perp(r_{\rm in})$.

These problems do not appear if we choose to have only an outer boundary, that is $r_{\rm in}=0$, in which case we have $\xi^\perp(r_{\rm in})=0$.  Then we can write 
 \ba\label{EffS2b}
 -\kappa  S^{(2)}_\lambda&\underset{\text{solu}}{=}&-\kappa  S^{(2)}_{\rm HJ}
 -\frac{1}{4} \int_{\partial M} \!\!\!\!\!d^{2} y \sqrt{h}
 \bigg[
 \rho  \Delta \rho\,- 2\rho \Delta   \xi^\perp(r_{\rm out}) +\xi^\perp(r_{\rm out}) \Delta  \xi^\perp(r_{\rm out})  
 \bigg] \,.\q\q \nn\\
 &=& 
 -\frac{1}{4} \int_{\partial M} \!\!\!\!\!d^{2} y \sqrt{h}
 \left(
 \rho \Delta \rho\,- 2\rho \, \delta \,{}^2\! R\right)
 -\frac{1}{4} \int_{\partial M} \!\!\!\!\!d^{2} y \sqrt{h} \xi^A {\cal D}_{AB} \xi^B   \q .
\ea
The $\rho$-dependent part is given by
\ba
S'_\rho&=&  -\frac{1}{4} \int_{\partial M} \!\!\!\!\!d^{2} y \sqrt{h}
 \left(
 \rho \Delta \rho\,- 2\rho \, \delta \,{}^2\! R\right)
\ea
and can be taken as dual boundary field theory, which reproduces the boundary--diffeomorphism invariant part of the gravitational Hamilton--Jacobi functional. 

Thus we see that for the case of a spherical boundary we produce exactly the action $S_\rho$ which we derived in section \ref{secGeod}, that is $S'_\rho=S_\rho$. Different from the cases with flat boundary discussed previously  there is no insertion of a non-local operator in $S'_\rho$.

Note that there are also special modes, that appear for the spherical boundary. Using spherical harmonics $Y^{lm}$ one will find that $\Delta$ is vanishing on $Y^{lm}$ with $l=1$. One thus has three modes $l=1$ and $m=-1,0,+1$ for which $S'_\rho$ is vanishing. These modes do describe the geometric position of the central point at $r=0$, which is encoded in the metric perturbations $\gamma_{AB}$ around $r=0$. Thus, we can understand these three modes as (diffeomorphism) gauge parameters for the gravitational field, which do happen to affect the geodesic length variable.

 As discussed above  we can use the Regge calculus set-up to argue that the geodesic effective action should be indeed local.  In (\ref{EffS2}) there is still the term $ \xi^A {\cal D}_{AB} \xi^B$, which is a priori non-local through the expressions of $\xi^A$ in terms of the boundary metric components $\gamma_{BC}$. Using the spherical (tensor) harmonics in Appendix \ref{AppHarm} one finds however that $\xi^A$ is determined by
 \ba
 \xi^\Psi=\tfrac{1}{2} \gamma^\Psi \q ,\q\q  \xi^\Phi=\tfrac{1}{2} \gamma^\Phi
 \ea
where we used an expansion $\gamma_{AB}= \gamma^\Psi \Psi_{AB}+ \gamma^\Phi \Phi_{AB}+ \gamma^\Theta \Theta_{AB}$ and $\xi_A=\xi^\Psi \Psi_A+ \xi^\Phi \Phi_B$ of the metric and vector field into tensor and vector harmonics respectively. Note that $\Psi_{AB}$ and $\Phi_{AB}$ are a basis for the trace free part of the metric perturbations. 

 \section{Discussion and outlook}\label{SecDisc}

In this work we determined holographic boundary theories for 3D linearized metric gravity, directly by computing the effective action for a geometric observable as determined from the gravitational action. This geometric observable is the geodesic distance from the boundary to some centre or central axis and describes so--called boundary degrees of freedom \cite{CarlipBreakingDiffeoAdS,FreidelDonnelly,Riello19}, or boundary gravitons. This degree of freedom encodes the shape of the (fluctuating) boundary in the embedding space time. Together with the holographic boundary theories we also determined the Hamilton--Jacobi functional for linearized gravity, for a large class of boundaries.

The resulting boundary theories depend on the chosen type of boundary and the choice of cosmological constant. It is known that Liouville theory arises for an asymptotic AdS boundary \cite{Henneaux, BarnichGonzales, CarlipBreakingDiffeoAdS,CotlerJensen}. We have shown that the effective theory for the geodesic lengths leads to Liouville like theories also for finite and more general boundaries. In particular one can always expect a Liouville-like coupling to the Ricci--scalar of the boundary. The reason is that the first variation of the Ricci--scalar is proportional to the first variation of the lengths of geodesics that start normal to the boundary. 

The boundary theories are furthermore defined by a quadratic form given by $\Delta=2(K^{CD}-Kh^{CD})D_CD_D-{}^2\!RK$. This gives a (non-degenerate) flat Laplacian for the torus boundary in AdS space and a degenerate Laplacian for the torus boundary in flat space. For a spherical boundary in flat space we obtain a differential operator proportional to the Laplacian on the sphere, but also a mass term resulting from ${}^2\!RK$. 

We have seen that in the case of a torus boundary  the derivation of the effective action for the geodesic lengths requires some subtle procedure. This is the imposition of smoothness condition at the central axis at $r=0$.  It leads to the insertion of a non-local operator $(1+\partial_\theta^{-2})$  into the effective action. 
 
 This has an important consequence, namely that the modes $k_\theta=\pm 1$ describe a gauge freedom of the boundary field theory. Indeed this follows from diffeomorphism symmetry modes, which affect the precise definition of the central axis. Accordingly the geodesic length at these modes is a gauge parameter and the geodesic effective action is independent of the boundary field and just given by the gravitational Hamilton--Jacobi functional, which does not depend on the boundary field, for $k_\theta=\pm 1$. The $k_\theta=0$ mode is also affected by diffeomorphism symmetry -- but here it is a diffeomorphism along the central axis, which to first order does not affect the lengths of the geodesics. In this case the geodesic effective action is given by the gravitational Hamilton--Jacobi functional, but with the addition of the Lagrange multiplier term, which imposes that the boundary field mode reproduces the geodesic length at $k_\theta=0$. 

This illustrates an interesting interplay between the bulk and the possibly asymptotic boundary. It deserves further study: for instance   the inclusion of a point particle at $r=0$ should change the smoothness conditions, and in fact break the gauge symmetry at $k_\theta=\pm 1$. Correspondingly one would expect that the one--loop partition function now reproduces a massive BMS character instead of the vacuum one, see also \cite{OblakThesis,DGLR2}. 

Another interesting direction is to investigate other geometric observables. For the asymptotically flat \cite{BarnichPhi} and AdS boundaries\cite{CotlerJensen} one can employ  certain angle variables, which are better suited to capture the BMS or Virasoro symmetry respectively. It would be interesting to see whether one can also identify germs for these symmetries at finite boundaries. It would also be interesting to study Lorentzian spacetimes, null boundaries and different boundary conditions \cite{CarlipBMS,Hopfi18,Wolf19}.

The  method to construct holographic duals directly from gravity, which we employed here, will allow us to study the 4D case. In this regard a first step has been taken in \cite{ADH}. Here a geodesic effective action has been computed for a 4D generalization of twisted thermal flat space.  For this  \cite{ADH} restricted to boundary conditions which impose flat perturbations, that is excluded propagating bulk gravitons. The resulting boundary action is then however a straightforward generalization of the 3D result, that is given by the same action with a Liouville like coupling to the boundary Ricci scalar and a degenerate kinetic term. The next step is to study how the inclusion of bulk gravitons affects the geodesic effective action, and in particular whether non-localities arise \cite{AsanteToappear}.  One might also be led to introduce additional boundary fields, which encode (better than the geodesic length) the dynamics of the bulk gravitons. A key question will be which kind of geometric observables are best suited for such boundary fields. 

We hope that these investigations will help for the understanding of the renormalization flow of quantum gravity models, e.g. \cite{AsymptoticSafety1,AsymptoticSafety2,AsymptoticSafety3}. A key issue  is to find suitable truncations, as one otherwise has to deal with an infinite dimensional space of possible couplings. The framework introduced in \cite{Dittrich12,TimeEvol, Dittrich14,Delcamp} employs boundaries and boundary Hilbert spaces to determine dynamically preferred truncation maps. Here a crucial question  is to identify geometric boundary observables which encode efficiently the bulk dynamics, which is also a key point in  the quasi-local holography program.

\begin{center}
\textbf{Acknowledgements }
\end{center}
BD would like to thank Sebastian Mizera and Aldo Riello for collaboration in the initial stages of the project.
FH is grateful for a Vanier Canada Graduate Scholarship.
SKA is supported by an NSERC grant awarded to BD. This work is  supported  by  Perimeter  Institute  for  Theoretical  Physics.   Research  at  Perimeter  Institute is supported by the Government of Canada through Industry Canada and by the Province of Ontario through the Ministry of Research and Innovation.

\appendix

  \section{Conventions and Gauss--Codazzi relations}\label{AppConv}
Here we collect some conventions for the curvature tensors and list the Gauss--Codazzi relations, which we make frequently use of.

The Riemann tensor is defined through the following commutator of covariant derivatives
\ba
(\nabla_a \nabla_b- \nabla_b \nabla_a) \xi_c &=& {R_{a b c}}^e \xi_e \q ,
\ea
and the Ricci tensor is given by $R_{ab}={R_{a cb}}^c$.
We define likewise the Riemann tensor for the boundary geometry, where we replace the space--time covariant derivative $\nabla_a$ with the spatial covariant derivative $D_A$. With our Gaussian coordinates we can define the extrinsic curvature as $K_{AB}= \tfrac{1}{2}\partial_\perp h_{AB}$.

For the class of maximally symmetric solutions, which we consider here,  the Riemann tensor  is given by
\ba\label{ConCurv1}
R_{abce} = \frac{2 \Lambda}{(d-1)(d-2)} \left( g_{a c} g_{be} - g_{ae} g_{bc} \right).
\ea

 The Gauss--Codazzi relation, which relates the Riemann tensor of the $d$--dimensional manifold $M$ and the Riemann tensor of the $(d-1)$--dimensional surfaces $r=\text{const.}$, states that 
\ba
^{(d-1)} {R_{ABC}}^D = {R_{AB C}}^D+ K_{AC} {K_{B}}^D - K_{BC} {K_{A}}^D \q ,
\ea

For vacuum solutions to the Einstein equations we have
\ba\label{BgEOM}
R_{ab} = \frac{2 \Lambda }{d-2} g_{ab} \q\q   \Rightarrow \q R=\frac{2d \Lambda}{d-2} 
\ea
and for such solutions the contracted Gauss--Codazzi relations become
\ba\label{GCVac1}
^{(d-1)}R_{AB} = \frac{2 \Lambda}{d-1} h_{AB} + K K_{AB} - {K_A}^C K_{CB} \q ,
\ea
\ba\label{GCVac2}
^{(d-1)}R = 2\Lambda + K^2 - K_{AB} K^{AB} \q .
\ea
The last equation coincides with the Hamiltonian constraint, that is the $(\perp\perp)$ component of the vacuum Einstein equations.

The Gauss--Codazzi relations furthermore state that
\ba\label{GCVac3}
D_A K_{BC}-D_B K_{AC} &=& R_{ABCe}n^e \,\underset{\text{max. sym. sol.}}{=} 0 \, , \nn\\
D_A {K_B~}^A-D_B{K_A~}^A &=& R_{Be} n^e\q \q\, \,\underset{\text{vac.-sol.}}{=} 0  \q .
\ea
The last set of relations $D_A {K_B~}^A-D_B{K_A~}^A=0$ coincide with the momentum constraints, that is the $(\perp A)$--components of the Einstein equations.

\section{Vector basis for induced perturbations}\label{AppB}

In this appendix, we show {\bf Result 1} relating  the diffeomorphism induced perturbations $\gamma_{AB}$ and the  components $\xi^\perp, \xi^A\p_A = \xi^\parallel$ of the diffeomorphism inducing vector field. The result holds for $2D$ boundaries with homogeneous scalar curvature on the background, $D_A ({}^2 R) = 0$, and a  3D background spacetime  satisfying the vacuum Einstein equations.

From \eqref{eq:gammaAB_in_xi} we see that a vector field $\xi^\perp \p_\perp + \xi^A \p_A$ acting on the background metric leads to an induced perturbation
\begin{align}\label{B1}
	\gamma_{AB} = 2 \xi^\perp K_{AB} + D_A \xi_B + D_B \xi_A,
\end{align}
where $K_{AB}, D_A$ and $\xi_A = h_{AB} \xi^B$ pertain to the background.

Our first claim is:\\
~\\
{\bf Claim:}
\begin{align}
	\Pi^{AB} \gamma_{AB} ={}& \Delta \xi^\perp \text{ where }\\
	\Pi^{AB} ={}& D^A D^B - D_C D^C h^{AB} - \frac12 ({}^2 R) h^{AB} \, , \\
	\Delta ={}& 2 (K^{CD} - K h^{CD}) D_C D_D - ({}^2 R) K.
\end{align}

~\\
{\bf Proof:}\\
 For any background and dimension we have
\begin{align}
(D^A D^B - D_C D^C h^{AB}) (D_A \xi_B + D_B \xi_A) ={}& D^B D^A D_B \xi_A + D^A D^B D_B \xi_A - 2 D_B D^B D^A \xi_A\nn\\
={}& (D^A D^B - D^B D^A) D_B \xi_A + 2 D^B (D_A D_B - D_B D_A) \xi^A\nn\\
={}& 2 D^B (R_{AB} \xi^A),
\end{align}
where we used that the first summand vanishes identically. For a 2D boundary, under the homogeneous curvature assumption, the last expression becomes $({}^2 \!R) D_A \xi^A$, hence
\begin{align}
	\Pi^{AB} (D_A \xi_B + D_B \xi_A) = 0.
\end{align}
Further note that
\begin{align}
	(D^A D^B - D_C D^C h^{AB}) (\xi^\perp K_{AB})={}& D_A \big(K^{AB} D_B \xi^\perp - K D^A \xi^\perp + \xi^\perp (D_B K^{AB} - D^A K) \big)\nonumber \\
	={}& K^{AB} D_A D_B \xi^\perp - K D_A D^A \xi^\perp + (D_A K^{AB}) D_B \xi^\perp - (D_A K) D^A \xi^\perp\nonumber \\
	={}& (K^{AB} D_A D_B - K D^A D_A) \xi^\perp,
\end{align}
where for the second and third lines we have used the momentum constraint  $D_A(K^{AB}-Kh^{AB})=0$  (which follows from the $(A\perp)$--components of the Einstein equations). Putting the previous two expressions together proves our first claim.

~\\
Secondly, we have the\\
~\\
{\bf Claim:}
\begin{align}
	\mathcal D^A{}_B \xi^B ={}& 2 (K^{BC} - K h^{BC}) \delta\,{}^2\Gamma^A_{BC} && \text{ where }\nonumber\\
	\mathcal D^A{}_B ={}& 2 (K^{CD} - K h^{CD}) D_C D_D h^A{}_B - ({}^2 \!R) K^A{}_B &&\text{ and }\nonumber\\
	\delta \, {}^2\! \Gamma^A_{BC} ={}& \frac12 h^{AD} (D_B \gamma_{CD} + D_C \gamma_{BD} - D_D \gamma_{BC}) &&
\end{align}
is the variation of the boundary Christoffel symbols.

~\\
To proof it we need a\\
{\bf Lemma:} Consider a $(d-1)$--dimensional hypersurface in a $d$--dimensional spacetime, which satisfies the vacuum Einstein solutions. We furthermore assume that the boundary comes with a Ricci tensor of the form ${}^{(d-1)}\! R_{AB}=\frac{1}{(d-1)} \, {}^{(d-1)}\! R h_{AB}$. Then we have:
\ba\label{KKIdentity}
(d-1) (K^A{}_C K^{CB} - K K^{AB}) = h^{AB} (K^{CD} K_{CD} - K^2) \q .
\ea
Note that for two--dimensional surfaces ${}^{2}\! R_{AB}=\frac{1}{2} \, {}^{2}\! R h_{AB}$ does hold automatically. This identity can be easily proven by using the Gauss--Codazzi relations for vacuum spacetimes (\ref{GCVac1}).

~\\
{\bf Proof:}
We start by writing
\begin{align}
	 2 (K^{BC} - K h^{BC}) \delta\,{}^2\!\Gamma^A_{BC} ={}& 2 (K^{BC}- K h^{BC}) D_B \gamma^A{}_C - (K^{BC}- K h^{BC}) D^A \gamma_{BC}.
\end{align}
where indices are raised with the induced background metric $h_{AB}$. Let us first evaluate the contribution of $\xi^\perp$, i.e., set $\gamma_{AB} = 2 \xi^\perp K_{AB}$ in the previous expression. One gets, distributing the derivatives,
\begin{align}
	&2 (K^{BC} - K h^{BC}) \delta\,{}^2\Gamma^A_{BC} \Big|_{\gamma_{AB} = 2 \xi^\perp K_{AB}} \nonumber\\
	={}& 4 (K^A{}_C K^{CB} - K K^{AB}) D_B \xi^\perp - 2 (K^{BC} K_{BC} - K^2) D^A \xi^\perp \nonumber \\
	& + \xi^\perp \Big( 4 (K^{BC} - K h^{BC}) D_B K^A{}_C - 2 (K^{BC} D^A K_{BC} - K D^A K) \Big).
\end{align}
The first line vanishes due to (\ref{KKIdentity}). 
For the first term in the second line we use the  momentum constraint and (\ref{KKIdentity}) to rewrite it as
\ba
4 (K^{BC} - K h^{BC}) D_B K^A{}_C \,=\, 2 D^A(K^{CD}K_{CD}-K^2) \q .
\ea
We have also for the second term in the second line
\ba
- 2 (K^{BC} D^A K_{BC} - K D^A K) \,=\, - D^A( K^{CD}K_{CD}-K^2) 
\ea
so that we remain with
\ba
2 (K^{BC} - K h^{BC}) \delta\,{}^2\Gamma^A_{BC} \Big|_{\gamma_{AB} }\,=\,  \xi^\perp  D^A(K^{CD}K_{CD}-K^2) \q .
\ea
The right hand side vanishes due to the Hamiltonian constraint (\ref{GCVac2}), which demands   that $K^2 - K_{CD} K^{CD}=({}^2 R) - 2 \Lambda$ and the homogeneous curvature assumption. 
Thus $\xi^\perp$ does not contribute to $2 (K^{BC} - K h^{BC}) \delta\,{}^2\Gamma^A_{BC}$. 

~\\
We are thus left with evaluating
\begin{align}
	2 (K^{BC} - K h^{BC}) \delta\,{}^2\Gamma^A_{BC} ={}& 2 (K^{BC} - K h^{BC}) \delta\,{}^2\Gamma^A_{BC} \Big|_{\gamma_{AB} = D_A \xi_B + D_B \xi_A}\nonumber\\
	={}& (K^{BC} - K h^{BC}) \big( (D_B D_C + D_C D_B) \xi^A +\, {}^2\!R_B{}^A{}_{CD} \xi^D +\, {}^2\!R_C{}^A{}_{BD} \xi^D \big)\nonumber\\
	={}& 2 (K^{BC} - K h^{BC}) D_B D_C \xi^A - R K^{AB} \xi_B,
\end{align}
where we have straightforwardly evaluated $\delta ^2\Gamma^A_{BC} \vert_{\gamma_{AB} = D_A \xi_B + D_B \xi_A}$ and used $^2\!R_{ABCD} = \frac12 ({}^2\! R) (h_{AC} h_{BD} - h_{AD} h_{BC})$. This proves the second claim.


\section{Second order of the Hamilton-Jacobi functional}\label{App2HJ}
Here we are going to prove:\\
~\\
{\bf Result 2:}  We consider a  2D boundary component $\partial M$ in a 3D space--time satisfying the vacuum Einstein equations. We assume the parametrization (\ref{B1}) for the boundary fluctuations $\gamma_{AB}$ in terms of the diffeomorphism generating vector field $\xi^a$. We furthermore assume that the boundary has homogeneous curvature $\partial_A {}^2\!R=0$. 
 
The second order of the Hamilton--Jacobi functional is then given by
 \ba\label{SHJ2App}
 -\kappa  S^{(2)}_{\rm HJ}&=& \frac{1}{4} \int_{\partial M} d^2 y \sqrt{h}  \epsilon \, \left(  \xi^\bot \Delta \, \xi^\bot \, -  \, \xi^A {\cal D}_{AB}  \xi^B  \right) \q ,
\ea
where 
\ba
\Delta&=& 2 (K^{CD} - K h^{CD}) D_C D_D - {}^2\! R \,K \q ,\nn\\
  {\cal D}_{AB} &=& \, 2 \left( K^{CD} -K h^{CD} \right)D_C D_D \,{h}_{AB} - {}^2\! R {K}_{AB} \q .
\ea
We remind the reader that we defined the extrinsic curvature tensor through the foliation, which with our choice of Gaussian coordinates amounts to $K_{AB}=\tfrac{1}{2}\partial_\perp h_{AB}$. We thus introduced $\epsilon $, which is equal to $+1$ for boundary components where   the outward pointing normal in the background geometry is given by $n \equiv \partial_\perp$ (that is the outer boundary), and $\epsilon =-1$ if $n \equiv -\partial_\perp$ (that is the inner boundary).
~\\~\\
{\bf Proof:}
We have to evaluate
\ba\label{AppCAct1}
-\kappa S^{(2)} &=&
\frac{1}{4}  \int_{\partial M} d^{2} y \, \epsilon \, \delta(\sqrt{h} \left(   K h^{AB}\  -  K^{AB}   \right)) \delta h_{AB}\,  .\q\q
\ea
with $\delta h_{AB}=\gamma_{AB}$ given by (\ref{B1}).

The following calculation applies to a space time of general dimension $d$, up to the point where we will explicitly set $d=3$ in (\ref{1.40}).
We abbreviate $\pi^{AB}= \sqrt{h}(K^{AB}-Kh^{AB})$ and find for the integrand in (\ref{AppCAct1})
 \ba
 {\cal F}&:=&\,\, \,\delta\left( \sqrt{h} \left(K h^{AB}-K^{AB}\right)\right) \delta h_{AB}\nn\\
 &=&-\delta\left(  \pi^{AB} \right)  \left( 2 \xi^\bot K_{AB} + {\cal L}_{\xi^\parallel} h_{AB} \right) \nn\\
&=& -\delta\left( \pi^{AB}  \left( 2 \xi^\bot K_{AB} + {\cal L}_{\xi^\parallel} h_{AB} \right) \right) + \pi^{AB} \delta\left( 2 \xi^\bot K_{AB}\right) +\pi^{AB} {\cal L}_{\xi^\parallel} \delta h_{AB} 
 \ea
 We use that by definition $\delta \xi^\bot =\delta \xi^A=0$ and that $ \pi^{AB} {\cal L}_{\xi^\parallel} h_{AB}$ is, modulo a total divergence, given by $2 \xi_B D_A\pi^{AB}$, where $D_A \pi^{AB}$ is the momentum constraint and thus vanishes. 
  Thus also the variation of $ \pi^{AB} {\cal L}_{\xi^\parallel} h_{AB}$ vanishes.  We again use $\delta h_{AB}=2\xi^\perp +  {\cal L}_{\xi^\parallel} h_{AB}$ in the last term, and indicate with $\simeq$ that we are calculating modulo total divergences:
 \ba
  {\cal F}&\simeq&-2 \xi^\bot  \delta\left( \pi^{AB}K_{AB}  \right) + 2 \xi^\bot \pi^{AB} \delta K_{AB} +  \pi^{AB} {\cal L}_{\xi^\parallel} (2 \xi^\bot K_{AB} )+ 
  \pi^{AB} {\cal L}_{\xi^\parallel} {\cal L}_{\xi^\parallel}  h_{AB} \, .\;\;
 \ea
We have for the third term
\ba
 \pi^{AB} {\cal L}_{\xi^\parallel} (2 \xi^\bot K_{AB} ) \,=\, {\cal L}_{\xi^\parallel} \left( \pi^{AB}  (2 \xi^\bot K_{AB} ) \right) 
  -2 \xi^\bot  {\cal L}_{\xi^\parallel} \left( \pi^{AB}  K_{AB} \right) + 2 \xi^\bot  \pi^{AB} {\cal L}_{\xi^\parallel} \left(  K_{AB} \right)
\ea
 and drop here the first term on the RHS, as it is a total derivative (since it is a Lie derivative of a scalar density). We find
 \ba\label{1.33}
   {\cal F}&\simeq& - 2 \xi^\bot \left( \delta   + {\cal L}_{\xi^\parallel} \right) \left( \pi^{AB}K_{AB}  \right)  + 2 \xi^\bot \pi^{AB}  \left( \delta   + {\cal L}_{\xi^\parallel} \right) K_{AB}   + 
  \pi^{AB} {\cal L}_{\xi^\parallel} {\cal L}_{\xi^\parallel}  h_{AB} \q .\q\q
 \ea
 Now for any derivation $\tilde \delta$
 \ba
\tilde \delta \left( \pi^{AB}K_{AB} \right)&=&\tilde \delta \left(   \sqrt{h}( h^{AC} h^{BD}-h^{AB} h^{CD}) K_{AB} K_{CD} \right) \nn\\
&=&
2 \sqrt{h}( h^{AC} h^{BD}-h^{AB} h^{CD}) K_{AB}\, \tilde\delta K_{CD} +   K_{AB} K_{CD}\, \tilde \delta \left(   \sqrt{h}( h^{AC} h^{BD}-h^{AB} h^{CD}) \right)
\nn\\
&=& 2 \pi^{AB} \, \tilde\delta K_{AB} + \left( \frac{1}{2}  \pi^{AB}K_{AB}   \, h^{CD} 
-2  \pi^{CA} {K_{A}}^D \right) \tilde  \delta h_{CD} \q .
 \ea
 
 Using  Lemma (\ref{KKIdentity}), which holds for boundaries with homogeneous curvature, we see that $(d-1)\pi^{CA} {K_{A}}^D\,=\,  \pi^{AB}K_{AB}   \, h^{CD}$ and we therefore have
 \ba
 \tilde \delta \left( \pi^{AB}K_{AB} \right)&=&2 \pi^{AB} \, \tilde\delta K_{AB} +\frac{(d-5)}{2(d-1)}  \pi^{AB}K_{AB}   \, h^{CD}\delta h_{CD}  \q .
 \ea
 We apply this identity for $-  \xi^\bot \left( \delta   + {\cal L}_{\xi^\parallel} \right) \left( \pi^{AB}K_{AB}  \right)$ in (\ref{1.33}) and obtain
 \ba
    {\cal F}&\simeq& -  \xi^\bot \left( \delta   + {\cal L}_{\xi^\parallel} \right)\!\! \left( \pi^{AB}K_{AB}  \right) - \tfrac{(d-5)}{2(d-1)} \xi^\perp \pi^{AB}K_{AB}   \, h^{CD} \left( \delta   + {\cal L}_{\xi^\parallel} \right)  h_{CD} + 
  \pi^{AB} {\cal L}_{\xi^\parallel} {\cal L}_{\xi^\parallel}  h_{AB} \nn\\
  &\simeq& -  \xi^\bot \left( \delta   + {\cal L}_{\xi^\parallel} \right)\! \!\left( K^{AB}K_{AB} -K^2 \right)  + 
    \tfrac{(3-d)}{(d-1)} \xi^\perp \pi^{AB}K_{AB}   \, h^{CD} \left( \delta   + {\cal L}_{\xi^\parallel} \right)\!\!  h_{CD}+
  \pi^{AB} {\cal L}_{\xi^\parallel} {\cal L}_{\xi^\parallel}  h_{AB}.\q\nn
 \ea

Now $K^{AB} K_{AB}-K^2\,=\,2 \Lambda - {}^{(d-1)}\!R  $ is the scalar constraint equation, which also holds  under the variation $\delta$. Furthermore, with our assumptions $^{(d-1)}\!R  -2 \Lambda$ is constant on the boundary and thus its Lie derivative vanishes. We remain with
\ba
  {\cal F}&\simeq&   \sqrt{h}  \xi^\bot  \delta  ^{(d-1)}\!R     
   +  \xi^\bot  \frac{(3-d)}{(d-1)} \sqrt{h}\left(   2 \Lambda -   ^{(d-1)}\!R   \right)             h^{CD}
  \left(  2 \xi^\bot K_{CD} +2 {\cal L}_{\xi^\parallel}  h_{CD}\right) \nn\\
 &&  \,  +\,
  \pi^{AB} {\cal L}_{\xi^\parallel} {\cal L}_{\xi^\parallel}  h_{AB} \, . \q\q
\ea
We now restrict to the $d=3$ and thus the second term on the right hand side vanishes. Using $ \delta  ^{2}\!R=\Delta \xi^\perp$ we obtain 
\ba\label{1.40}
 {\cal F}&\underset{d=3}{\simeq}&   \sqrt{h}  \xi^\bot \Delta \xi^\bot \, +  \pi^{AB} {\cal L}_{\xi^\parallel} {\cal L}_{\xi^\parallel}  h_{AB} \, . 
\ea

For the last term in (\ref{1.40}) we can write
\begin{align}
 \pi^{AB} \L_{\xi^\parallel} \L_{\xi^\parallel} h_{AB} & =  2 \pi^{AB} \L_{\xi^\parallel} (D_A \xi_B)\nn\\
 & =  2 \pi^{AB} \left(\xi^C D_C D_A \xi_B + D_A \xi_C D_B \xi^C + D_C \xi_B D_A \xi^C\right)\nn\\
 & = -2 \pi^{AB} \left(- \xi^C D_C D_A \xi_B + \xi_C D_A D_B \xi^C + \xi^C D_A D_C \xi_B\right) \,\nn\\
 &\q +\left[ 2 \pi^{AB} \left(  D_A (\xi_C D_B \xi^C )+ D_A( \xi^C  D_C \xi_B \right) \right]\nn\\
 & \simeq - 2 \xi^C \pi^{AB} D_A D_B \xi_C - 2 \pi^{AB} \, {}^2\!R_{ACBD} \xi^C \xi^D \q .
\end{align}
 where going from the third to the fourth equation we dropped the term in square bracket, as it is a total divergence due to the momentum constraint $D_A{\pi^A}_B=0$.
 
Finally ${}^2\!R_{ACBD} = \frac12  {}^2\! R (h_{AB} h_{CD} - h_{AD} h_{CB})$, giving. 
\begin{align}
 \pi^{AB} \L_{\xi^\parallel} \L_{\xi^\parallel} h_{AB}  \,
&\simeq\, -2 \xi^C \, \pi^{AB} D_A D_B \xi_C +\sqrt h \xi^C  \, {}^2\!R K \, \xi_C + \, {}^2\!R\, \pi^{AB}  \xi_A \xi_B\nn\\
& = - \sqrt{h} \, \xi^A  {\cal D}_{AB} \xi^B  \q .
\end{align}
 
Thus we have for the 3D Hamilton--Jacobi functional
\ba
-\kappa S^{(2)}_{HJ} &=&
\frac{1}{4}  \int_{\partial M} d^{2} y 
   \sqrt{h}  \epsilon \,  \left( \xi^\bot \Delta \xi^\bot \, -  \, \xi^A {\cal D}_{AB} \xi^B \right) \q .
\ea

\section{Evaluation of the commutator $\left[\partial_\perp,\Delta\right]$}\label{AppD}
 
 For Appendices \ref{AppE} and \ref{AppF} we will need the evaluation of the commutator
 \ba
 \left[\partial_\perp, \Delta^{-1}\right]&=& -\Delta^{-1} \left[ \partial_\perp,\Delta\right] \Delta^{-1} \q .
 \ea
With 
$ \Delta=2(K^{CD}-Kh^{CD})D_CD_D-{}^2\! R K$
let us therefore consider
\ba
\partial_\perp \Delta f
&=&
\Delta \partial_\perp f
+2 \left(\partial_\perp(K^{CD}-Kh^{CD})\right) D_C D_D f \nn\\&&
- 2(K^{CD}-Kh^{CD}) (\partial_\perp \,\, {}^2\! \Gamma^E_{CD}) D_E f 
-(\partial_\perp ( ^2\! R)) Kf - ^2\! R (\partial_\perp K) f 
\ea
where $f$ is a scalar function.
We  compute all the terms appearing in this expression. To start with we employ the Ricci equation adapted to Gaussian coordinates 
 \ba
 {R_{A\perp B}}^\perp\,&=& \partial_\perp \Gamma^\perp_{AB}-\sum_C \Gamma^C_{\perp B} \Gamma^\perp_{C A} \nn\\
 &=& -\partial_\perp K_{AB} \,+\, K^C_BK_{CA}
 \ea
 With $R_{A\perp B\perp}=\Lambda g_{AB}$ we obtain
 \ba
 \partial_\perp K_{AB} &=& K_A^C K_{BC} - \Lambda h_{AB} 
 \ea
 We also have $\p_\perp h_{AB} = 2 K_{AB}$ and $\p_\perp h^{AB} = -2K^{AB} $. Therefore
  \ba
 \partial_\perp K^{AB} &=& -3K^{AC} K^B_{C} - \Lambda h^{AB}\q ,\nn\\
  \p_\perp K &=& - K_{AB}K^{AB} - 2\Lambda \q , \nn \\ 
 \partial_\perp (K_{AB} K^{AB}) &=&  - 2K^{A}_{C}K^{CB} K_{AB} - 2 \Lambda K \q .
 \ea
To find the radial derivative of the Ricci scalar, we apply the  Gauss--Codazzi relation for the Ricci scalar (before taking the radial derivative) and for the Ricci tensor (after taking the radial derivative). This gives
\ba
\p_\perp ( {}^2\!R ) \underset{\text{G.-C.}}=& - 2 \, ({}^2\!R^{AB} )K_{AB} = -{}^2\!RK
\ea
where we used that ${}^2\!R^{AB}=\tfrac{1}{2}\, {}^2\!R h^{AB}$.
%
Thus, using the Gauss--Codazzi relations repeatedly
\ba
\p_\perp (K^{CD} - K h^{CD}) &=& K_{AB}K^{AB} h^{CD} + \Lambda h^{CD} + 2 K K^{CD} -3K^{DE}K^{C}_{E} \nn \\
&\underset{\text{G.-C.}}=&  K^2 h^{CD}   - K K^{CD} + {}^2\!R^{CD} 
\ea
For the radial derivative of the Christoffel symbols we compute
\ba\label{prChris}
\partial_\perp ({}^2\!\Gamma^E_{CD}) &=& \tfrac{1}{2} h^{EF} \left( D_C \partial_\perp h_{DF}+ D_D \partial_\perp h_{CF}-D_F \partial_\perp h_{CD}\right)\nn\\
&=&
h^{EF}\left(D_C K_{DF} + D_D K_{CF} - D_F K_{CD}\right) \nn\\
&\underset{\text{G.-C.}}=& D_C K_{D}^E\,\,=\,\, D_D K_{C}^E \,\,=\,\, D^E K_{CD}
\ea
Therefore we have to consider the term
\ba
-2 (K^{CD}-Kh^{CD}) (\partial_\perp \,\, {}^2\!\Gamma^E_{CD}) D_E f &\underset{\text{G.-C.}}=& 2 (Kh^{CD}-K^{CD}) (D_CK^E_{D}) D_E f \nn \\
&\underset{\text{G.-C.}}=& 2D_C  (KK^{CE}-K^{CD}K^E_{D} ) D_E f \nn \\
&\underset{\text{G.-C.}}=& 2D_C  ( {}^2\!R^{CE}- \Lambda h^{CE} ) D_E f \q .
\ea
This term vanishes due to our homogenous curvature condition $D_A {}^2\!R= 0$. 
Hence we have 
\ba\label{Commuf1}
\partial_\perp \Delta f
&=&
\Delta \partial_\perp f + 2(K^2 h^{CD}   - K K^{CD} +  {}^2\!R^{CD} )D_C D_D f +  {}^2\!R K^2 \,    f + {}^2\!R(2\Lambda + K_{CD}K^{CD} )f \nn \\
&=& \Delta \partial_\perp f - K \Delta f + 2\, {}^2\!R^{CD} D_C D_D f + {}^2\!R(2\Lambda + K_{CD}K^{CD} )f  \nn \\
&=& \Delta \partial_\perp f - K \Delta f + {}^2\!R( h^{CD} D_C D_D + 2\Lambda + K_{CD}K^{CD} )f  \q ,
\ea
and thus
\ba\label{Commuf2}
 \left[ \partial_\perp,\Delta\right] &=& - K \Delta  + {}^2\!R( h^{CD} D_C D_D + 2\Lambda + K_{CD}K^{CD} ) \q ,\label{Comm1} \\
 \left[\partial_\perp, \Delta^{-1}\right]&=& \Delta^{-1} \left(     K \Delta  - {}^2\!R( h^{CD} D_C D_D + 2\Lambda + K_{CD}K^{CD} )             \right) \Delta^{-1} \q .  \label{Comm2}
\ea

 \section{Solutions of the equations of motion for the case ${}^2\!R=0$}\label{AppE}

 The second order gravitational action with Lagrange multiplier term (\ref{ActionLambda}) is given by
\ba\label{ActionLambdaApp}
-\kappa  S^{(2)}_\lambda
&=& 
\frac{1}{4} \int_M d^3 x  \sqrt{g} \, \,\gamma_{a b} \left(   
  V^{abcd} \,
  \gamma_{cd} \,\,+ \,\,
  \tfrac{1}{2}  \,
  G^{abcdef }
   \, \nabla_c \nabla_d \gamma_{ef}  \right) \,+ \nn\\
   &&
  \frac{1}{4} \int_{\partial M} d^{2} y   \sqrt{h}\,\epsilon \,
\gamma_{ab}  \left(\,(B_1)^{abcd}\gamma_{cd} +   \, (B_2)^{abcde} \nabla_c \gamma_{de} \right)  \,+\nn\\
 && \frac{1}{4} \int_{(\partial M)_{\rm out}} d^{2} y\, \, \lambda(y) \left( \rho(y) - \ell[ \gamma_{\perp\perp}] \right)
\ea
where  
\ba
V^{abcd} &=& \frac{1}{2} \left[ \frac{1}{2}\left(R-2\Lambda \right)  \left( g^{ab}  g^{cd} - 2 g^{ac} g^{bd} \right) - R^{ab}  g^{cd}   -  g^{ab}  R^{cd} + 2\left( g^{ac} R^{bd}+ g^{bc} R^{ad}   \right) \right] \\ 
G^{ab ef cd} &=&  g^{ab} g^{ec}g^{fd} +g^{ac} g^{bd}g^{ef} + g^{ae} g^{bf}g^{cd} - g^{ab} g^{ef}g^{cd} - g^{af} g^{bd}g^{ec} - g^{ac} g^{bf}g^{ed}    \\
B_1^{abcd} &=&  \frac{1}{2} ( K h^{ab}-K^{ab} ) g^{cd} - h^{ac} h^{bd}K - h^{ab}K^{cd} + h^{ac} K^{bd}+ h^{bc}K^{ad}     \\ 
B_2^{ab e cd} &=&  \frac{1}{2} \left( \left(  h^{ae} h^{bd} - h^{ab} h^{ed} \right)n^c + \left(  h^{ac} h^{be} - h^{ab} h^{ce} \right) n^d -  \left(  h^{ac} h^{bd} - h^{ab} h^{cd} \right)n^e \right). \q 
\ea
The derivation for the second order expansion of the boundary term can be found in \cite{BonzomDittrich}.

Using the form of the Ricci tensor and Ricci scalar for vacuum solutions, we can write 
\ba
V^{abcd} = \frac{\Lambda}{d-2}  \left( 2 g^{ac} g^{bd} - g^{ab}  g^{cd}  \right) \q .
\ea
The variation of the action (\ref{ActionLambdaApp}) with respect to the metric perturbations $\gamma_{ab}$ leads to the equations of motion
 \ba\label{EOMlambdaApp}
\hat G^{ab}:=\left(   
  V^{abcd} \,
  \gamma_{cd} \,\,+ \,\,
  \tfrac{1}{2}  \,
  G^{abcdef }
   \, \nabla_c \nabla_d \gamma_{ef}  \right)&=& \frac{1}{4} \frac{\lambda(y)}{\sqrt{h}} \delta^{a}_\perp \delta^b_\perp   \q .
\ea
One can show that $\gamma_{ab}=\nabla_a\xi_b+\nabla_b \xi_a$ satisfies (\ref{EOMlambdaApp}) for $\lambda=0$. Here we want to solve the equations including the Lagrange multiplier term. As explained in section \ref{BigSecEffGL} it is sufficient to solve the $(\perp\perp)$ and $(\perp A)$ components for of the equations of motion for the lapse and shift perturbations.

In the following we will therefore consider the Hamiltonian constraint
\ba\label{Hdef}
H\,:=\, 2V^{\perp\perp cd} \,
  \gamma_{cd} \,\,+ \,\,
  G^{\perp\perp cdef } \, \nabla_c \nabla_d \gamma_{ef}  \q ,
\ea
as well as the momentum constraint
\ba\label{Mdef}
M^A\,:=\,2 V^{\perp A cd} \,
  \gamma_{cd} \,\,+ \,\,
  G^{\perp A cdef } \, \nabla_c \nabla_d \gamma_{ef}  \q .
\ea

To rewrite the constraints we will make use of the fact that we have a maximally symmetric background solution (\ref{ConCurv1}) and that the Gauss--Codazzi relations (\ref{GCVac1},\ref{GCVac2},\ref{GCVac3}) hold. 

Using our Gaussian coordinates we furthermore replace the space-time covariant derivatives with
\ba
\nabla_A \gamma_{\perp\perp} &=& D_A \gamma_{\perp\perp} - 2K_{A}^B \gamma_{\perp B}
  ,\nn  \\
\nabla_A \gamma_{\perp B} &=& D_A \gamma_{\perp B} - K_{A}^C \gamma_{B C} + K_{AB} \gamma_{\perp \perp},\nn \\
\nabla_A \gamma_{BC} &=& D_A \gamma_{BC} + K_{AB} \gamma_{\perp  C} + K_{AC} \gamma_{\perp B},\nn \\
\nabla_\perp \gamma_{AB} &=& \partial_\perp \gamma_{AB} - K^E_A\gamma_{EB} - K^E_B\gamma_{AE },\nn\\
\nabla_\perp \gamma_{\perp B} &=& \partial_\perp \gamma_{A\perp} -   K^E_A\gamma_{E\perp} ,
\ea
where the spatial covariant derivative $D_A$ acts on only the spatial indices $B$. Note that $\nabla_\perp \gamma_{AB}$ involves only the spatial metric perturbations.

Employing the equations above we  can expand the following expressions quadratic in the covariant derivatives:
\ba
\nabla_A \nabla_B \gamma_{CD} &=& D_A D_B \gamma_{CD} + D_A \left( K_{BC} \gamma_{\perp D} + K_{BD} \gamma_{\perp C } \right) + K_{AC} D_B \gamma_{\perp D} + K_{AD} D_B \gamma_{\perp C}  \q\nn \\&&\, + K_{AB} \nabla_\perp  \gamma_{CD} 
 + (K_{AC} K_{BD} + K_{AD}K_{BC}) \gamma_{\perp \perp } - K_{AC} K_B^E \gamma_{DE} - K_{AD} K_B^E \gamma_{CE} ,\nn \\ 
\nabla_D \nabla_C \gamma_{\perp B} &=& D_D D_C \gamma_{\perp B} + D_D \left( K_{BC} \gamma_{\perp \perp } \right) - D_D(K_C^E \gamma_{BE}) + K_{CD} \nabla_\perp  \gamma_{\perp B} + K_{BD}D_C \gamma_{\perp \perp }  \nn \\&& \,
 \q-2 K_C^E K_{BD}\gamma_{\perp E}  - K_D^E \left( D_C\gamma_{BE}+  K_{BC}\gamma_{\perp E} +K_{CE}\gamma_{\perp B}  \right),\nn \\
 \nabla_B \nabla_\perp  \gamma_{CD}  &=& D_B \nabla_\perp  \gamma_{CD} + K_{BC} \nabla_\perp  \gamma_{\perp D} + K_{BD} \nabla_\perp  \gamma_{\perp C}  \nn \\&& \,
\,\,\q- K_{B}^E \left( D_E\gamma_{CD} + K_{CE} \gamma_{\perp D} + K_{DE} \gamma_{\perp C} \ \right) \q .
\ea
With these ingredients the Hamiltonian and momentum constraints become after some algebra
\ba
H
&=& 2(K h^{AB} - K^{AB} )  D_A \gamma_{\perp B} + (  {}^{2}\!R - 2\Lambda ) \gamma_{\perp \perp } -\Lambda h^{AB}\gamma_{AB} \nn \\
&&\,\, -   H^{ABCD} \left( D_A D_B \gamma_{CD} + K_{AB}\nabla_\perp  \gamma_{CD}  - K_{AC}K_B^E \gamma_{DE} \right)  \,\, , \\
M^A 
&=&  2 ( {}^{2}\!R^{AB})\gamma_{\perp B} + (K^{AB}-Kh^{AB})D_B \gamma_{\perp \perp } + H^{ABCD} (D_D D_C\gamma_{\perp B} + D_B \p_\perp  \gamma_{CD} ) \nn \\
&& -H^{ABCD} (D_D(K_C^E\gamma_{BE} ) +D_B(K_C^E\gamma_{DE}) + D_B(K_D^E\gamma_{CE})  + K_B^ED_E \gamma_{CD} + K_D^E D_C \gamma_{BE} ) \q  \q \nn\\
\ea
where we abbreviated $H^{ABCD} := h^{AB}h^{CD} - h^{AC}h^{BD}$.

To commute the radial derivative with the spatial covariant derivatives, we employ (\ref{prChris})
\ba
\p_\perp D_B \gamma_{CD} &=& D_B \p_\perp \gamma_{CD} - (\p_\perp {}^{(d-1)}\!\Gamma^E_{BC})\gamma_{DE} - (\p_\perp {}^{(d-1)}\!\Gamma^E_{BD})\gamma_{DE} \nn \\
&\underset{(\ref{prChris})}=& D_B \p_\perp \gamma_{CD}  - D_B K^E_C \gamma_{DE} - D_B K^E_D \gamma_{CE} \q .\q
\ea
Thus the momentum constraint can be simplified to 
\ba\label{Mom15}
M^A &=&  2( {}^{2}\!R^{AB})\gamma_{\perp B} + (K^{AB}-Kh^{AB})D_B \gamma_{\perp \perp } + H^{ABCD} (D_D D_C\gamma_{\perp B} + \p_\perp  D_B \gamma_{CD} ) \nn \\
&& -H^{ABCD} ( D_D(K_C^E\gamma_{BE} )+K_C^ED_B\gamma_{DE} +K_D^ED_B\gamma_{CE}+ K_B^ED_E \gamma_{CD} + K_D^E D_C \gamma_{BE} ) \, . \q\q\q
\ea

We will now restrict to the case ${}^{2}\!R_{AB}=0$. Note that terms of the form
\ba
H^{ABCD} D_A D_D  { T^{\cdots}}_{\cdots} \,=\, \left(D^BD^C-D^CD^B\right){ T^{\cdots}}_{\cdots} 
\ea
are now vanishing as they involve the boundary curvature tensor.  Furthermore $H^{ABCD}D_AK_D^E = 0$ as $D_AK_D^E=D_DK_A^E=D^EK_{AD}$ due to the Gauss--Codazzi relation (\ref{GCVac3}).
We thus obtain for the divergence of the momentum constraint
\ba
D_AM^A &=&   D_A((K^{AB}-Kh^{AB})D_B \gamma_{\perp \perp} )+ H^{ABCD} (D_A\p_\perp D_B \gamma_{CD} ) \nn \\
&& -H^{ABCD} D_A( K_C^ED_B\gamma_{DE} +K_D^ED_B\gamma_{CE}+ K_B^ED_E \gamma_{CD} + K_D^E D_C \gamma_{BE} ) \q\q \nn \\
&=&   (K^{AB}-Kh^{AB})D_AD_B \gamma_{\perp\perp} +\p_\perp ( H^{ABCD}  D_A D_B \gamma_{CD} ) -( \p_\perp H^{ABCD})  D_A D_B \gamma_{CD} \, \nn \\
&& -H^{ABCD} ( K_C^ED_AD_B\gamma_{DE} +K_D^ED_AD_B\gamma_{CE}+ K_B^ED_AD_E \gamma_{CD} + K_D^E D_AD_C \gamma_{BE} ) \, .\q\q
\ea
where we have used the Gauss Codazzi relations, the fact that $H^{ABCD}D_AK_D^E = 0$ and the formula for commuting the radial derivative with the spatial covariant derivative to arrive at the second equation.

Now with $\p_\perp h^{AB} = -2 K^{AB}$ we have for the radial derivative of $H^{ABCD}=h^{AB}h^{CD} - h^{AC}h^{BD}$
\ba \label{radialH}
(\p_\perp H^{ABCD})D_A D_B \gamma_{CD} &=&-2 (h^{AB}K^{CD} + h^{CD}K^{AB} - h^{AC}K^{BD} - h^{BD}K^{AC} ) D_A D_B \gamma_{CD}\nn \\
&=&- 2H^{ABCD} ( K_C^ED_AD_B\gamma_{DE} +K_D^ED_AD_B\gamma_{CE} \nn\\
&&\q\q\q\,\, \,+ K_B^ED_AD_E \gamma_{CD} + K_D^E D_AD_C \gamma_{BE} ) \q. \q
\ea
Therefore, the momentum constraint is further simplified as
\ba
D_AM^A &=&  (K^{AB}-Kh^{AB})D_AD_B \gamma_{\perp \perp} +\p_\perp ( H^{ABCD}  D_A D_B \gamma_{CD} )- \tfrac{1}{2}\p_\perp (H^{ABCD})  D_A D_B \gamma_{CD} \nn\\
&=& \tfrac{1}{2}\Delta \gamma_{\perp \perp} -\p_\perp ( \Pi^{AB} \gamma_{CD} )- \tfrac{1}{2}\p_\perp (H^{ABCD})  D_A D_B \gamma_{CD}\q .
\ea
where we have used the definitions $\Delta=2(K^{AB}-Kh^{AB})D_AD_B$ and $\Pi^{AB}\gamma_{AB}=-H^{ABCD}D_AD_B\gamma_{CD}$, which apply to the case that ${}^2\!R=0$.

We now adopt the parametrization
\ba
\gamma_{CD} = 2 \xi^\perp K_{CD} + D_C \xi_D + D_C \xi_D
\ea
for the boundary metric. We note that
\ba
(\p_\perp H^{ABCD})D_A D_B (D_C \xi_D + D_C \xi_D)&=&0
\ea
as we have ${}^2\!R=0$ and thus can commute the spatial covariant derivatives. Furthermore, using again that spatial covariant derivatives commute we have
\ba
(\p_\perp H^{ABCD})D_A D_B (2K_{CD} \xi^\perp)&=&
-4K^{CD}h^{AB}
 \big( D_A D_B (K_{CD} \xi^\perp) - D_AD_C (K_{BD} \xi^\perp) \nn\\
 && \q\q\q\q\q \;+D_C D_D (K_{AB} \xi^\perp) -D_C D_A (K_{DB} \xi^\perp) 
\big) \, .\q
\ea
Employing the Gauss--Codazzi relation (\ref{GCVac3}) repeatedly one sees that all the derivatives acting on the extrinsic curvature tensor cancel out and we are left with
\ba
(\p_\perp H^{ABCD})D_A D_B (2K_{CD} \xi^\perp)&=&
-4 \left( K^{CD}K_{CD} h^{AB} +K K^{AB}-2 K^{AC}{K_{C}}^B\right)D_AD_B \xi^\perp\nn\\
&\underset{(\ref{KKIdentity})}{=}&4 K \left( K^{AB}-K h^{AB}\right)D_AD_B \xi^\perp \nn\\ 
&=& 2K \Delta \xi^\perp  \q .
\ea
We thus arrive at
\ba
D_AM^A &=&
 \tfrac{1}{2}\Delta \gamma_{\perp \perp} -\p_\perp \Delta \xi^\perp - K \Delta \xi^\perp \nn\\
 &=& \tfrac{1}{2}\Delta \gamma_{\perp \perp} -\Delta \p_\perp  \xi^\perp 
\ea
where for the last equation we have used the result (\ref{Commuf1}) for the commutator of $\Delta$ and $\partial_\perp$.  

Therefore, in the case that ${}^2\!R=0$ the solution for the lapse is given by 
\ba
\gamma_{\perp\perp}=2\partial_\perp \xi^\perp \q , 
\ea
even if we include the Lagrange multiplier term into the equation of motion, as in (\ref{EOMlambdaApp}).

We now insert this solution into  the Hamiltonian 
\ba
H
&=& 2(K h^{AB} - K^{AB} )  D_A \gamma_{\perp B}  - 2\Lambda  \gamma_{\perp \perp } -\Lambda h^{AB}\gamma_{AB} \nn \\
&&\,\, -   H^{ABCD} \left( D_A D_B \gamma_{CD} + K_{AB}\nabla_\perp  \gamma_{CD}  - K_{AC}K_B^E \gamma_{DE} \right) \q .
\ea
Inserting furthermore $\gamma_{\perp A}=D_A\xi^\perp+h_{AB} \p_\perp \xi^B$, and making use of the identities we have derived so far, one finds that $H=0$.  To take into account the Lagrange multiplier term consider the ansatz $\gamma_{\perp A}=D_A\xi^\perp+h_{AB} \p_\perp \xi^B + t_{B}$. We then have to solve
\ba
H=2(K h^{AB} - K^{AB} )  D_A t_B \,\stackrel{!}{=}\, \frac{1}{2} \frac{\lambda}{\sqrt{h}} \q .
\ea 
Choosing $t_B =-D_B \frac{1}{2\Delta}  \frac{\lambda}{\sqrt{h}} $ we see that we satisfy this equation. The addition of such a $t_B$ to $\gamma_{\perp B}$ does also leave the momentum constraint (\ref{Mom15}) invariant, if we use that ${}^2\!R=0$ and spatial covariant derivatives commute. 

Thus we find that in the case of vanishing spatial curvature the Lagrange multiplier term is accommodated by the solutions
\ba
\gamma_{\perp\perp} &=& 2\p_\perp \left( \xi^\perp - \frac{1}{2\Delta}  \frac{\lambda}{\sqrt{h}}  \right)\,=\,  2\p_\perp  \xi^\perp \, ,\nn\\
\gamma_{\perp A}&=& D_A(\xi^\perp  - \frac{1}{2\Delta}  \frac{\lambda}{\sqrt{h}} )+h_{AB} \p_\perp \xi^B
\ea
where the second equation in the first line follows from the commutator of $\partial_\perp$ and $\Delta$ (for ${}^2\! R=0$) in (\ref{Commuf1}) and the fact that $\partial_\perp h^{-1/2}=-Kh^{-1/2}$.

\section{Equations of motion  in spherical coordinates}\label{AppSph}

Here we will consider the equations of motions with Lagrange multiplier term (\ref{EOMlambdaApp}) for the case with spherical boundary, where we have ${}^2\!R\neq 0$. Using the definitions for the Hamiltonian $H$ and momentum constraints $M^A$ in (\ref{Hdef}) and (\ref{Mdef}) respectively, as well as the expressions (\ref{Dsph}) for the operators $\Delta$ and ${\cal D}$, one can verify in a straightforward way that the following equations hold for the background metric (\ref{sphMetric}):
\ba\label{E1110}
D_A M^{A} - \frac{1}{r}  \!\left(H-\frac{1}{2}  \frac{\lambda}{\sqrt{h}} \right) 
&=& \frac{1}{2} \Delta \gamma_{\perp\perp} - \frac{1}{r^3} \partial_\perp\left(  r^3 \Pi^{AB} \gamma_{AB} \right)  + \frac{1}{2r} \frac{\lambda}{\sqrt{h}} \label{E1110}\\
D_A \!\left(H-\frac{1}{2}\frac{\lambda}{\sqrt{h}}  \right) +  \frac{2}{r} M_A
&=& -{{\cal D}_A}^B \gamma_{\perp B} +D_A\! \left( \Pi^{CD} \gamma_{CD}\right)      -\frac{1}{2} D_A  \frac{\lambda}{\sqrt{h}}  -\frac{h_{AB}}{r^3} \partial_\perp\! \left( 2 r^2 h^{CD} \delta {}^2\!\Gamma^B_{CD}\right)  \,.  \label{E1111}   \q\q
\ea

For the geometry of the sphere we have 
\ba
\Delta\,=\, \frac{1}{r^3} \tilde \Delta
\ea
where we define $\tilde \Delta$ to be equal to $\Delta$, but with the radial coordinate set to $r=1$. Thus $ \tilde \Delta \partial_\perp \,=\,  \partial_\perp  \tilde \Delta$ and we can rewrite (\ref{E1110}) as
\ba\label{E1110b}
D_A M^A - \frac{1}{r} \left( H-\frac{1}{2}\lambda\right)&=& \frac{1}{2} \Delta \gamma_{\perp\perp} - \Delta \partial_\perp \Delta^{-1}  \left(  \Pi^{AB} \gamma_{AB} \right) +\frac{1}{2r}\frac{\lambda}{\sqrt{h}}  \q .
\ea

To treat the remaining equations we  observe that ${{\cal D}^A}_B =\tfrac{1}{r^3} {\tilde{\cal D}^A}_B$ and thus
\ba
\frac{1}{r^3} \partial_\perp \, r^3  V^A &=& \left( { \cal D} \partial_\perp {\cal D}^{-1} V \right)^A  \q 
\ea
for any vector field $V^A$.
Furthermore, remember that both ${\cal D}$ and $\Delta$ contain a  Laplacian operator $D^C D_C$. For the commutation of $D^C D_C$ with $D^A$ we have
\ba
D^C D_C D^A f \,=\,    D^A D^C D_C f \,+\,       {}^2\!R^{AE} D_E f \,=\,    D^A D^C D_C f \,+\,    \tfrac{1}{2}   {}^2\!R h^{AE} D_E f \q .
\ea
for any scalar $f$. Using that $D^A\,  {}^2\!R=0$ and $D^A  K=0$ for the sphere we see that
\ba
{\cal D}^{AB} D_B f\,=\, D^A \Delta f  \q .
\ea

This allows us to write (\ref{E1111}) as 
\ba\label{E1111b}
D_A \left(H -\frac{1}{2} \frac{\lambda}{\sqrt{h}} \right)+  \frac{2}{r} M_A
&=& -{{\cal D}_A}^B \gamma_{\perp B} +    {{\cal D}_A}^B   D_B \Delta^{-1} \left( \Pi^{CD} \gamma_{CD}\right) -\frac{1}{2}  {{\cal D}_A}^B   D_B \Delta^{-1}  \frac{\lambda}{\sqrt{h}}
 \nn\\ &&+ {\cal D}_{AB}  \partial_\perp   \left( {\cal D}^{-1} \left(  - \frac{2}{r}h^{CD} \delta {}^2\!\Gamma^\cdot_{CD}\right) \right)^B\q\q
\ea

In summary we obtain the solutions
\ba
\gamma_{\perp\perp} &=& 2 \partial_\perp \hat\xi^\perp   \q , \nn\\
\gamma_{\perp A}
&=& D_A \hat \xi^\perp  \,+\,  h_{AB}\partial_\perp   \xi^B   
\ea
 with
 \ba
 \hat \xi^\perp \,=\, \xi^\perp -\frac{1}{2\Delta} \frac{\lambda}{\sqrt{h}}
 \ea
and $\xi^\perp$ and $\xi^A$ defined in (\ref{xiofg}).  These solutions can be inserted into the spatial--spatial part of the Einstein equations, and one will find that these evaluate to zero, $\hat G^{AB}=0$. This can be also expected from the fact that the divergence of the Einstein equations, including the Lagrange multiplier term, vanishes identically.

 \section{Lagrange multiplier dependent boundary terms}\label{AppF}


For our examples we consider the Einstein equations with Lagrange multiplier term
\ba
\hat G^{ab}&=& \frac{1}{4} \frac{\lambda(y)}{\sqrt{h}} \delta^{a}_\perp \delta^b_\perp   \q ,
\ea
and solve the $G^{\perp\perp}$ and $G^{\perp A}$ equations for the lapse and shift perturbations $\gamma_{\perp\perp}$ and $\gamma_{AB}$. In all cases we find that the solutions can be expressed as
\ba\label{AppSolLS}
\gamma_{\perp\perp} &=& 2 \partial_\perp \hat \xi^\perp   \q , \nn\\
\gamma_{\perp A}
&=& D_A \hat \xi^\perp  \,+\,  h_{AB}\partial_\perp   \xi^B    \q ,
\ea
where 
\ba
\hat \xi^\perp \,=\, \xi^\perp-\frac{1}{2\Delta} \frac{\lambda}{\sqrt{h}}
\ea
and the components $\xi^a$ are understood as functionals of the spatial metric perturbations $\gamma_{AB}$, as defined in (\ref{xiofg}). In particular we see that the addition of the Lagrange multiplier term results in the shift of $\xi^\perp$ to $\hat \xi^\perp$. 

Here we are going to evaluate the (second order) boundary term on solutions of the form (\ref{AppSolLS}). We already know the result for $\lambda$=0 (see Appendix \ref{App2HJ}), we therefore need only keep track of the $\lambda$--dependent terms. 

These terms only arise through the lapse and shift components. The only terms where these appear in the second order contribution to the boundary action (\ref{ActionLambdaApp}) are given by
\ba
&& \gamma_{AB} \left( B_1^{AB\perp\perp} \gamma_{\perp\perp} + B_2^{ABC \perp D} \nabla_C \gamma_{\perp D} + B_2^{ABC C \perp} \nabla_D \gamma_{C\perp} \right)\nn\\
&=& \gamma_{AB} \left( 
\tfrac{1}{2}(Kh^{AB}-K^{AB})\gamma_{\perp\perp} + (h^{AC}h^{BD}-h^{AB}h^{CD})  \nabla_C \gamma_{\perp D}
\right)
\ea
For the covariant derivative of the shift components we have
\ba
\nabla_C \gamma_{\perp D} &=& D_C  \gamma_{\perp D} - \Gamma^E_{C\perp} \gamma_{ED}- \Gamma^\perp_{CD} \gamma_{\perp \perp}\nn\\
&\underset{\gamma_{AB}=0}{=}& D_C  \gamma_{\perp D} + n^\perp K_{CD} \gamma_{\perp \perp} \nn\\
&\underset{\gamma_{AB}=0}{=}&- \frac{1}{2} D_C D_D  \frac{1}{\Delta} \frac{\lambda}{\sqrt{h}} - n^\perp K_{CD}\, \partial_\perp \frac{1}{\Delta} \frac{\lambda}{\sqrt{h}} 
\ea
and for the lapse  
\ba
\gamma_{\perp\perp}&=& -\partial_\perp \frac{1}{\Delta} \frac{\lambda}{\sqrt{h}}  \q .
\ea
Thus we obtain
\ba
&&
\left( 
\tfrac{1}{2}(Kh^{AB}-K^{AB})\gamma_{\perp\perp} + (h^{AC}h^{BD}-h^{AB}h^{CD})  \nabla_C \gamma_{\perp D}
\right)\nn\\
&\underset{\gamma_{AB}=0}{=}&
-\tfrac{1}{2}  \left(D^AD^B-h^{AB} D^C D_C\right) \frac{1}{\Delta} \frac{\lambda}{\sqrt{h}} - \tfrac{1}{2} \left( K^{AB}-Kh^{AB}\right) \partial_\perp \left(\frac{1}{\Delta} \frac{\lambda}{\sqrt{h}}\right) \q .
\ea
 Now, with (\ref{Comm2}) we have
 \ba
 \partial_\perp \left(\frac{1}{\Delta} \frac{\lambda}{\sqrt{h}}\right) &=& \frac{1}{\Delta} \partial_\perp \frac{\lambda}{\sqrt{h}}  +  \Delta^{-1} \left(     K \Delta  - {}^2\!R( h^{CD} D_C D_D + 2\Lambda + K_{CD}K^{CD} )             \right) \Delta^{-1}  \frac{\lambda}{\sqrt{h}}  \nn\\
 &=& -\Delta^{-1} \,\,   {}^2\!R( h^{CD} D_C D_D + 2\Lambda + K_{CD}K^{CD} )         \Delta^{-1}  \frac{\lambda}{\sqrt{h}} 
 \ea
 where we used $\partial_\perp h^{-1/2}=-h^{-1/2}K$ and $\partial_\perp\lambda=0$.
 ~\\
 Let us first consider the case that ${}^2\!R=0$, which applies to the cases in sections \ref{SecFlat} and \ref{SecAdS}. In this case we have that $\Pi^{AB}=D^AD^B-D^CD_Ch^{AB}$. We then obtain for the $\lambda$--dependent boundary term
 \ba
&&-  \frac{1}{8} \int_{\partial M} d^{2}y \,  \sqrt{h}\epsilon \,  \gamma_{ab} \left(\left(D^AD^B-h^{AB} D^C D_C\right) \frac{1}{\Delta} \frac{\lambda}{\sqrt{h}} \right) \nn\\
&=& -  \frac{1}{8} \int_{\partial M} d^{2}y \,  \sqrt{h} \epsilon \, \left( \Delta^{-1} \Pi^{AB} \gamma_{AB} \right)\frac{\lambda}{\sqrt{h}}\nn\\
&=&-  \frac{1}{8} \int_{\partial M} d^{2}y\,\epsilon \, \xi^\perp \q .
\ea
 
Secondly we have for the spherical boundary embedded in flat space, that is the case in section \ref{SecSph}, $K=\tfrac{2}{r}$ and $K_{AB}=\tfrac{1}{2}Kh_{AB}$ as well as ${}^2\!R=\tfrac{2}{r^2}$ so that $\Delta=-\tfrac{2 }{r}\left(D^C D_C+\tfrac{2}{r^2}\right)$. This gives
\ba
\left( K^{AB}-Kh^{AB}\right) \partial_\perp \left(\frac{1}{\Delta} \frac{\lambda}{\sqrt{h}}\right) 
 &=&\!\!  -\left( K^{AB}-Kh^{AB}\right)\frac{1}{\Delta}    {}^2\!R( h^{CD} D_C D_D + 2\Lambda + K_{CD}K^{CD} )       \frac{1}{\Delta}\frac{\lambda}{\sqrt{h}} \nn\\
 &=&\frac{1}{r} h^{AB}\frac{1}{\Delta}  \frac{2}{r^2}\left(D^C D_C+\frac{2}{r^2}\right) \frac{1}{\Delta} \frac{\lambda}{\sqrt{h}} \nn\\
 &=& -\frac{1}{r} h^{AB} \frac{1}{\Delta}  \frac{1}{r}  \Delta \frac{1}{\Delta}  \frac{\lambda}{\sqrt{h}} \nn\\
 &=&-\frac{1}{2} \, {}^2\!R h^{AB} \frac{1}{\Delta}  \frac{\lambda}{\sqrt{h}} \q .
\ea
 With $\Pi^{AB}=D^AD^B-D^CD_Ch^{AB} -\tfrac{1}{2}{}^2\!Rh^{AB}$ we obtain that also in this case, the $\lambda$--dependent terms in the boundary term are given by
 \ba
-  \frac{1}{8} \int_{\partial M} d^{2}y \,  \sqrt{h} \epsilon\, \left( \Delta^{-1} \Pi^{AB} \gamma_{AB} \right)\frac{\lambda}{\sqrt{h}}
&=&-  \frac{1}{8} \int_{\partial M} d^{2}y\,\epsilon \,  \xi^\perp \lambda \q . 
 \ea
  
\section{Geodesic length to first order in metric perturbations}\label{AppG}

We are interested in the geodesic distance between two fixed coordinate points, for a given (Euclidean) metric. The full metric will differ from a  background metric by a perturbation, and we need the expansion of the geodesic distance in the metric perturbations to first order.

The background metric is $g_{ab} $, and the background geodesic $x^a(\tau)$ with $\tau \in [0,1]$. The full metric is $g^{\rm full}_{ab}$ and the $g^{\rm full}$-geodesic will be called $z(\tau)=x(\tau)+\delta z(\tau)$. We will assume that $z$ is affinely parametrized, so it has constant modulus w.r.t $g^{\rm full}$ and thus satisfies
\begin{align}
 \nabla^{\rm full}_{\dot z} \dot z^a = 0, \qquad \frac{d}{d\tau} (\dot z^a \dot z^b g^{\rm full}_{a b}) = 0.
\end{align} 
These equations continue to hold under variations.

We will consider the variation of the square of the geodesic length
\begin{align}
 L^2 = \int_0^1 d \tau \,  \dot z^a \dot z^b g^{\rm full}_{ab} \q .
\end{align}
This is indeed the square length as $z(\tau)$ has constant modulus. The variation is given by
\begin{align}
 \delta L^2 =& \int_0^1 \d\tau \Big[ 2 \frac{d}{d\tau} \big( \delta z^a \dot x^b g_{ab} \big) - 2 \delta z^a\frac{d}{d\tau} \big( g_{ab} \dot x^b \big) + \delta z^c \p_c g_{ab} \dot x^a \dot x^b + \dot x^a \dot x^b \gamma_{ab}\Big]\\
 =& 2 \Big[\delta z^a \dot x^b g_{ab} \Big]_{\tau = 0}^1 - 2 \int_0^1 d \tau \, \delta z^a (\nabla_{\dot x} \dot x_a) + \int_0^1 \d \tau \dot x^a \dot x^b \delta \gamma_{ab} \q ,
\end{align}
where second and third term in the first line combine to the covariant derivative in the second line.

Since $\delta z$  vanishes at the end points, we can drop the first term. Furthermore as $x$ is an affine geodesic, we can also drop the second term and are left with
\begin{align}
\delta L^2 = \int_0^1 d \tau \, \dot x^a \dot x^b \gamma_{ab} \q .
\end{align}

With $x^a=(r_{\rm in}+(r_{\rm out}-r_{\rm in})\tau,0,0)$ we therefore have for the first order perturbation of the geodesic length
\begin{align}
 \ell:= \delta L = \frac{1}{2(r_{\rm out}-r_{\rm in})}    \int_0^1 d \tau \, (r_{\rm out}-r_{\rm in})^2 \, \gamma_{\perp\perp}\,=\,\frac{1}{2} \int_{r_{\rm in}}^{r_{\rm out}}  d \tau \, \gamma_{\perp\perp}(r) \q .
\end{align}

\section{Smoothness conditions for the metric at $r=0$}\label{AppSmooth}

Consider metric perturbations $\gamma_{\mu\nu}$ expressed in Cartesian coordinates  $(x,y,t)$ such that the components of the metric $\gamma_{\mu\nu}$ are smooth at the origin and can thus be expanded in a Taylor series in the coordinates.  We shall transform the metric from flat into polar coordinates $(r,\theta, t)$ and spherical coordinates $(r,\theta, \varphi)$ and  study the behaviour of the metric components near the origin $r \rightarrow 0$. Let us denote the components of the metric perturbations  in polar or spherical coordinates by $\gamma_{ab}$. 

In polar coordinates, we have the transformation of the coordinates and the components of the metric are given by
\ba
x = r \cos \theta, \q y = r \sin \theta, \q \gamma_{ab} = \gamma_{\mu\nu} \frac{\p x^{\mu}}{\p x^a}\frac{\p x^{\nu}}{\p x^b} \q.
\ea
The components of the metric in polar coordinates are therefore given by 
\ba
\gamma_{\perp\perp} &=& \gamma_{xx}\cos^2 \theta  + \gamma_{yy} \sin^2 \theta + \gamma_{xy}\sin2\theta, \nn \\
\gamma_{\theta \theta} &=& r^2 \left( \gamma_{xx}\sin^2 \theta  + \gamma_{yy}\cos^2 \theta -   \gamma_{xy}\sin2\theta  \right), \nn \\
\gamma_{t t} &=& \gamma_{tt} ,\nn\\
\gamma_{\perp \theta} &=& r \left( \tfrac{1}{2} \sin(2\theta) ( \gamma_{yy} - \gamma_{xx} ) + \gamma_{xy}  \cos2\theta  \right), \nn \\
\gamma_{\perp t} &=&\gamma_{xt}  \cos \theta + \gamma_{yt}  \sin\theta ,\nn\\
\gamma_{\theta t} &=& r \left( \gamma_{yt}\cos \theta  - \gamma_{xt}\sin \theta  \right)  \q.
\ea
Given that the metric components $(\gamma_{xx},\gamma_{yy}, \gamma_{tt},\gamma_{xy},\gamma_{xt},\gamma_{yt})$ are smooth functions near the origin, a Taylor expansion of the metric perturbations around the origin for the thermal flat spinning space is given by
\ba
\gamma_{ab} &=& a_{ab}^{(0)} +  a_{ab}^{(1)} \, r +a_{ab}^{(2)} \, r^2 + {\cal O}(r^3)  \q \text{for } ab = rr, tt, rt ; \nn \\
\gamma_{ab} &=&  \q \,\, a_{ab}^{(1)} \, r + a_{ab}^{(2)} \, r^2 + {\cal O}(r^3)  \q \text{for } ab = r\theta, \theta t ;\nn \\
\gamma_{\theta \theta} &=&  \q \q \q a_{\theta \theta}^{(2)} \, r^2 + {\cal O}(r^3) \q .
\ea

In spherical coordinates, we have the coordinate transformation 
\ba
x = r \sin \theta \cos \varphi, \q y = r \sin \theta \sin \varphi , \q t = r \cos \theta  \q.
\ea
The components of the metric in spherical coordinates are given by 
\ba
\gamma_{\perp\perp} &=& \sin^2\! \theta(  \gamma_{xx} \cos^2\!\varphi +\gamma_{yy} \sin^2 \!\varphi  ) + \gamma_{tt}\cos^2\!\theta  + \gamma_{xy}\sin^2\!\theta \sin\!2\varphi  + \sin\!2\theta (\gamma_{yt}\sin\!\varphi  + \gamma_{xt}\cos\!\varphi  ) ,\nn \\
\gamma_{\theta \theta} &=& r^2 \left( \cos^2\! \theta(  \gamma_{xx} \cos^2\!\varphi +\gamma_{yy} \sin^2 \!\varphi  ) +\gamma_{tt}\sin^2\!\theta  + \gamma_{xy}\cos^2\!\theta \sin\!2\varphi  - \sin\!2\theta (\gamma_{yt}\sin\!\varphi  + \gamma_{xt}\cos\!\varphi  ) \right) ,\nn \\
\gamma_{\varphi  \varphi } &=& r^2 \left( \sin^2 \!\theta(  \gamma_{xx} \sin^2\!\varphi +\gamma_{yy} \cos^2 \!\varphi  ) - \gamma_{xy}\sin^2\!\theta \sin\!2\varphi    \right)  ,\nn\\
\gamma_{\perp \theta} &=& r\! \left( \tfrac{1}{2} \sin\!2\theta ( \gamma_{xx} \cos^2\!\varphi +\gamma_{yy} \sin^2\! \varphi  - \gamma_{tt} + \gamma_{xy}(\sin\!2\varphi  + \cos\!2\theta)  ) \!    -\!2\sin^2\!\theta (\gamma_{xt}\cos\!\varphi  +\gamma_{yt}\sin\!\varphi )  \right) , \nn \\
\gamma_{\perp \varphi } &=&  r \sin^2\! \theta \left( \tfrac{1}{2}\sin\!2\varphi  (  \gamma_{yy}  - \gamma_{xx}  ) +\gamma_{xy}\cos^2\! \varphi  \right), \nn\\
\gamma_{\theta \varphi } &=& r^2\sin\!2\theta \left( \tfrac{1}{4}\sin\! 2\varphi  (\gamma_{yy} - \gamma_{xx} ) + \gamma_{xy} \cos^2\!\varphi   \right)  \q.
\ea
The Taylor expansion for the metric perturbations in spherical spacetime region around the origin $r=0$ is thus
\ba
\gamma_{\perp\perp} &=& a_{rr}^{(0)} +  a_{rr}^{(1)} \, r +a_{rr}^{(2)} \, r^2 + {\cal O}(r^3) ;\nn \\
\gamma_{ab} &=&  \q \,\, a_{ab}^{(1)} \, r + a_{ab}^{(2)} \, r^2 + {\cal O}(r^3)  \q \text{for } ab = r\theta, r\varphi  ;\nn \\
\gamma_{ab} &=&  \q \q \q a_{ab}^{(2)} \, r^2 + {\cal O}(r^3)  \q \text{for } ab = \theta\theta, \varphi  \varphi  ,  \theta \varphi  \q . 
\ea

\section{On effective actions }\label{AppEff}

Here we will consider a quadratic dynamical system with two dynamical variables $(x,y)$ and integrate out one of these variables $y$ in order to define an effective action for the remaining variable $x$. We will then consider the case that the action for the variable $y$ is degenerate and show that the effective action will take a special form. This can be easily generalized to systems with more variables.

We start with an action
\ba
S_\lambda=\frac{1}{2} \left( \begin{array}{c} x \\ y \end{array}\right)^{\rm t} \cdot M \cdot \left( \begin{array}{c} x \\ y \end{array}\right)\,+\,
\left( \begin{array}{c} x \\ y \end{array}\right)^{\rm t} \cdot \left( \begin{array}{c} b_x \\ b_y \end{array}\right) \,+\, 
\left( \begin{array}{c} (\rho-x) \\ 0 \end{array}\right)^{\rm t} \cdot \left( \begin{array}{c} \lambda \\ 0 \end{array}\right) 
\ea
with dynamical variables $(x,y)$, ``boundary values" $(b_x,b_y)$ and a Lagrange multiplier term, which enforces $x=\rho$. We will assume that the matrix $M$ is invertible.

Variation with respect to $x$ and $y$ leads to equations of motion, which are solved by
\ba
\left( \begin{array}{c} x \\ y \end{array}\right)&=& -M^{-1} \cdot  \left( \begin{array}{c} b_x \\ b_y \end{array}\right) + M^{-1} \cdot \left( \begin{array}{c} \lambda \\ 0 \end{array}\right)  \q .
\ea
We will now differentiate two cases, firstly the case $(i)$ $M_{yy}\neq 0$ (or in the higher--dimensional case $\text{det}M_{yy}\neq 0$) and secondly the case $(ii)$, which is that $M_{yy}=0$. \\
~\\
In case $(i)$,
as
\ba
(M^{-1})_{xx} \,=\, \frac{M_{yy}}{\text{det}M}
\ea
we find that the solution for $x$ is $\lambda$--dependent. Let us denote by $x_0[b_x,b_y]$ the solution for $\lambda=0$. Then we have the solution
\ba
x= x_0[b_x,b_y] + \frac{ M_{yy}}{\text{det}M} \lambda \q ,
\ea
which we insert into the Lagrange multiplier equation $\rho=x$ and solve for $\lambda$:
\ba
 \lambda  =\frac{\text{det}M}{M_{yy}}  \left(  \rho-x_0[b_x,b_y] \right) \q .
\ea
Inserting these solutions back into the action we find
\ba\label{J6}
S_\lambda &\underset{{\rm sol. \,for}\, x,y}{=}& 
-\frac{1}{2} \left( \begin{array}{c} b_x \\ b_y \end{array}\right)^{\rm t} \cdot M^{-1}  \cdot \left( \begin{array}{c} b_x \\ b_y \end{array}\right)+
\frac{1}{2} \left( \begin{array}{c} \lambda \\ 0 \end{array}\right)^{\rm t} \cdot M^{-1}  \cdot \left( \begin{array}{c} \lambda \\ 0 \end{array}\right)
 +\lambda(\rho-x_0[b_x,b_y] - \tfrac{M_{yy}}{\text{det}M} \lambda)
  \nn\\
&\underset{{\rm sol. \,for}\, \lambda}{=}&
\frac{\text{det}M}{M_{yy}}\left(\frac{1}{2} \rho^2  -\rho \,x_0[b_x,b_y]\right) +\frac{1}{2}\frac{\text{det}M}{M_{yy}} (x_0[b_x,b_y] )^2
\,-\,\frac{1}{2} \left( \begin{array}{c} b_x \\ b_y \end{array}\right)^{\rm t} \cdot M^{-1}  \cdot \left( \begin{array}{c} b_x \\ b_y \end{array}\right)\, ,\q
\ea
which can be adopted as effective action for the dynamical variable $\rho=x$.

~\\
For case $(ii)$ we will however find that the solution for $x$ does not depend on $\lambda$, but is  determined only by  the boundary values $x= x_0[b_x,b_y]$. Thus we {\it cannot} solve the Lagrange multiplier equation $\rho=x=x_0[b_x,b_y]$ for $\lambda$. We have rather to understand this equation as a condition on the parameter $\rho$. Evaluating the action on the solution we obtain 
\ba
S_\lambda &\underset{{\rm sol. \,for}\, x,y}{=}& 
-\frac{1}{2} \left( \begin{array}{c} b_x \\ b_y \end{array}\right)^{\rm t} \cdot M^{-1}  \cdot \left( \begin{array}{c} b_x \\ b_y \end{array}\right)+
 \lambda(\rho-x_0[b_x,b_y] )
\ea
where $\lambda$ remains a free variable, enforcing $\rho= x_0[b_x,b_y]$. The term quadratic in $\lambda$ which appears in (\ref{J6}) is now vanishing, as we have $(M^{-1})_{xx}=0$. Thus the on-shell action is just given by the on-shell action of $S_{\lambda=0}$ plus the Lagrange multiplier term, with the solution for $x$ inserted.

\section{Spherical tensor harmonics}\label{AppHarm}

Here  we define scalar, vector and tensor spherical harmonics.  These spherical harmonics are eigenfunctions of the Laplace operator and are furthermore characterized by their divergence and their trace \cite{Sandberg}. We denote by $Y^{lm}$  the scalar spherical harmonics  (and omit the indices $(l,m)$). Furthermore we consider here a unit sphere, that is fix $r=1$. The vector and tensor harmonics are defined by
\ba
&&\Psi_A= D_A Y, \q\q\q\q \q\q\q\q\q\q\q
\Phi_A = { \epsilon_A\,}^B  D_B Y,   \nn\\
&&\Psi_{AB}= D_B D_A Y + \tfrac{1}{2} l(l+1)  h_{AB} Y ,\q
\Phi_{AB}\,=\, \tfrac{1}{2}\left( D_A \Phi_B+ D_B \Phi_A\right)  ,\q\q
\Theta_{AB}=  h_{AB} Y  \q .\q
\ea
with
${  \epsilon_\theta ~ }^\varphi=\sin^{-1} \!\theta$ and ${  \epsilon_\varphi~ }^\theta = -\sin \theta$.

We have the following properties for $\square=D^AD_A$:
\ba
&&\square Y \,=\, -  l(l+1) Y \, , \nn\\
&&\square \Psi_B \,=\, (1-l(l+1)) \Psi_A \, ,\q  \q \square \Phi_B \,=\, (1-l(l+1)) \Phi_A \, ,   \nn\\
&&\square \Psi_{BC} \,=\, (4-l(l+1)) \Psi_{BC}  ,\q  \square \Phi_{BC} \,=\, (4-l(l+1)) \Phi_{BC}  \, , \q
\square \Theta_{BC} \,=\,-l(l+1) \Theta_{BC}  \, .\q\q
\ea
Furthermore
\ba
&&D^A\Psi_A= - l(l+1) Y  \, , \; \q\q\q\q D^A \Phi_A\,=\, 0 \, \q \nn\\
&&D^A \Psi_{AB}= \tfrac{1}{2}(2-l(l+1))\Psi_B  ,\q D^A \Phi_{AB}\,=\,  \tfrac{1}{2}(2-l(l+1))\Phi_B ,\q
D^A \Theta_{AB} = \Psi_B \, .\q\q
\ea
Finally we have for the trace of the tensor modes
\ba
h^{AB} \Psi_{AB}\,=\, 0  \, ,\q h^{AB} \Phi_{AB}\,=\, 0  \, , \q h^{AB} \Theta_{AB}\,=\, 2 Y    \q .
\ea

\end{document}